\documentclass[aps,prb,reprint]{revtex4-1}
\usepackage{graphicx}
\usepackage{dcolumn}
\usepackage{bm}
\usepackage{hyperref}
\usepackage{amssymb}
\usepackage{amsmath}
\usepackage{epsf}
\usepackage[caption=false]{subfig}
\usepackage{epstopdf}
\usepackage{amsfonts}
\usepackage{color}

\usepackage{pst-all}
\usepackage{graphicx}
\usepackage[section]{placeins}
\usepackage{natbib}
\begin{document}
\title{Controlling local currents in molecular junctions}
\author{Hari Kumar Yadalam and Upendra Harbola}
\affiliation{Department of Inorganic and Physical Chemistry, Indian Institute of
Science, Bangalore, 560012, India.}
\begin{abstract}
The effect of non-equilibrium constraints and dephasing on the circulating
currents in molecular junctions are analyzed. Circulating currents are 
manifestations of quantum effects  
and can be induced either by externally applied bias or an external magnetic
field through the molecular system. In symmetric Aharonov-Bohm ring, bond currents have 
two contributions, bias driven 
and magnetic field driven. We analyze the competition between these two
contributions and show that, as a consequence, current through one of the
branches can be completely suppressed. We 
then study the effect of asymmetry (as a result of chemical substitution) on 
the current pathways inside the molecule and study asymmetry induced circulating currents
(without magnetic field) by tuning the coupling strength of the substituent (at finite bias). 
\end{abstract}
\maketitle
\section{Introduction}
Persistent charge current is the current flowing in systems with ring 
geometries, 
due to the phase coherent motion of electrons \cite{imry}. It can be induced by 
the presence of vector 
potential due to magnetic fields threading the ring, first studied by Pauling in
the context of aromatic molecules \cite{pauling}. 
The effect of magnetic flux through superconducting discs was studied by
Byers and Yang \cite{byers}. Buttiker et.al. \cite{buttiker}, showed that 1D 
metallic rings, where phase coherence of electrons 
is maintained, act like superconductors to produce a persistent current flowing
in the ring. Further studies have explored the effects of various quantities 
like
disorder, temperature and coulomb interaction between electrons on the 
persistent current \cite{imry}. 
In this study we extend these works to the molecular regime in 
the molecular junction setup.\par
\indent
In recent years, the electron conduction through
single molecular junction has attracted a lot of research interest due to its
fundamental interest in exploring 
quantum effects and its applications in miniaturization of electronic components
(molecular electronics).  
The idea of molecular electronics is to control the electronic current by
manipulating the physical and chemical properties of the molecule. Current
flowing through molecules in junction can take 
different pathways inside a molecule. These pathways have been 
studied recently \cite{solomon}. 
Here we analyze the local currents inside a molecular junction in 
presence (in symmetric junction) and absence (asymmetric junction) of the
magnetic field.\par
\indent Magnetic field threading the molecular 
ring induces different phases in the electron wavefunction as it 
transverses through different pathways inside the molecule.
As we discuss below, this phase acquired by the electron affects both 
the local currents and the net current. Hence the dependence of current flow on 
magnetic flux allows to control not only the net current through the junction but 
also the local bond currents inside the junction. Although much work has been done to study
the effect of magnetic flux on current flowing between leads 
\cite{rai,solomon,Hod1,Hod2}, 
little attention has been paid to the study of the effect of magnetic flux on
the circulating currents inside the molecule in 
the presence of external bias, with notable exceptions of references 
\cite{Rai1,Rai2,Jayannavar1,Davidovich}.
Here we analyze the aspect of controlling the local currents by manipulating 
external magnetic field and chemical substitution. For a symmetric 
Aharonov-Bohm ring case, in the presence of the external bias, it is possible to 
fine tune the magnetic flux to completely suppress 
current flow across different branches selectively. 
We show that the bond current has two contributions, which we
identify as magnetic field driven and bias driven contributions. These two 
contributions compete and may cancel each other along a branch, while add up to
enhance the current along the other branch. This is not possible if either the 
magnetic field or the bias is present alone.
 We further consider the case where
an extra site is coupled to the ring system and demonstrate that a circulating 
current (in this work we adapt an intuitive definition of circulating current as, "circulating current 
is present if the direction of current flowing through one of the branches is opposite to the 
direction of the net current, and its magnitude is given by the smallest of the currents flowing 
across the two branches") 
can also be induced by tuning the coupling strength of the substituent (at finite bias). 
This is due to the asymmetry induced 
between pathways by the extra coupling site. We derive analytic expressions for 
the bond currents and discuss them under different conditions. We find that 
the circulating currents can be induced not only by the 
magnetic field but also due to coupling with the leads (in the presence of asymmetry and finite bias). That is 
the direction of the current flowing across a branch can be manipulated by tuning the coupling strength 
with the leads. We present a detailed analysis of bond currents based on 
analytical results.\par
\indent The rest of the paper is organized as follows. In the next section 
(Sec.(\ref{gen})) we consider a model with asymmetry in the presence of magnetic 
field and calculate the bond currents inside the molecule and the net current in the 
circuit.
In Sec, (\ref{sym-abring}), we present a symmetric molecular ring system 
coupled to two metal leads in the 
presence of a magnetic flux. We discuss bond currents, net current and 
the circulating current at equilibrium  (when the two leads are at the same 
thermodynamic state) and 
non-equilibrium conditions. In Sec. (\ref{chem-subs}) we discuss circulating
currents in an asymmetric molecular ring junction in 
absence of the magnetic field. We conclude in Sec. (\ref{con}).
\section{Model Hamiltonian and Current calculations}
\label{gen}
\subsection*{Model Hamiltonian}
\begin{figure}
\centering
\includegraphics[width=7.0cm,height=3.8cm]{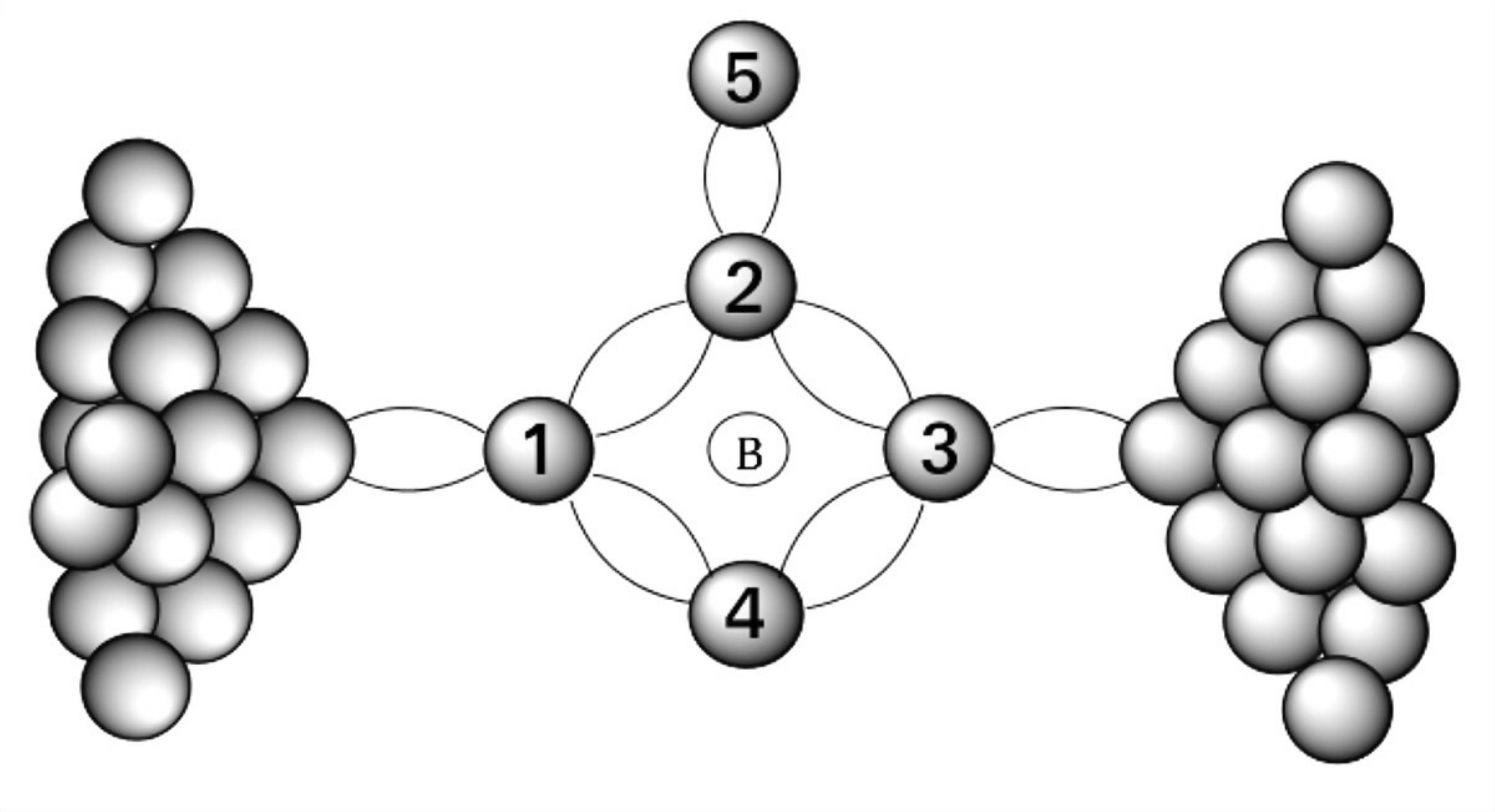}
\caption{(Color online) Schematic of model system considered. It consists of 
four identical localized 
sites coupled to each other to form a ring geometry (with magnetic field 
piercing the ring), 
diagonally opposite sites are coupled to two metal leads and one of the sites 
not coupled to leads is coupled to an extra site.}
\label{model_g}
\end{figure}
To study the effect of asymmetry and magnetic fields on bond currents in ring 
molecular systems out of equilibrium (at steady-state), we consider a simple 
model shown in Fig. \ref{model_g}. 
It consists of a ring molecular system with four identical localized sites 
(orbitals) coupled  to nearest sites through hopping. Diagonally 
opposite sites are coupled to two metal leads, 
and one of the sites not coupled to leads is coupled to an extra site. Further 
a 
magnetic flux is pierced through the molecular ring. 
All four sites forming the ring are taken to have the same energy (taken as 
zero 
by rescaling all other energies) and their coupling strengths to nearest 
neighbors are 
equal, taken as energy unit. Specifically, sites '1' $\&$ '3' 
are coupled to left and right metallic leads (modeled as free electron 
reservoirs at thermal equilibrium) respectively. 
Site '2' is coupled to an extra site (site '5' with energy '$\epsilon$') with 
coupling strength '$t$'. Effect of the external magnetic field is included in 
the model Hamiltonian in the spirit of 
Peierls substitution \cite{peierls}.\par
\indent
The Hamiltonian describing this model is given as, 
\begin{eqnarray}
\label{eq-1}
 \hat{H}&=&\sum_{i,j=1}^5 H_{0_{ij}}^{} c_i^{\dag} 
c_j^{}+\mathop{\sum_k}_{\alpha=L,R}\epsilon_{\alpha,k}^{} d_{\alpha k}^\dag 
d_{\alpha k}^{}\nonumber\\
         &+&\sum_k\big[g_L^{} d_{Lk}^\dag c_1^{}+g_R^{} d_{Rk}^\dag 
c_3^{}+h.c.\big]
\end{eqnarray}
where
\begin{align}
\label{eq-2}
 H_{0}^{}=\begin{pmatrix}
          0&-e^{-i\frac{\phi}{4}}&0&-e^{i\frac{\phi}{4}}&0\\
          -e^{i\frac{\phi}{4}}&0&-e^{-i\frac{\phi}{4}}&0&-t\\
          0&-e^{i\frac{\phi}{4}}&0&-e^{-i\frac{\phi}{4}}&0\\
         -e^{-i\frac{\phi}{4}}&0&-e^{i\frac{\phi}{4}}&0&0\\
          0&-t&0&0&\epsilon\\
         \end{pmatrix}
\end{align}
is the single particle Hamiltonian for the isolated molecule. '$\phi$' is the 
dimensionless magnetic flux given by $(B \times A)/(\frac{\hbar c}{e})$, where 
'B' is the strength of applied magnetic field, 
'A' is the area of the molecular ring and $\hbar$, c, e represents reduced 
Planck's constant, speed of light and (absolute) charge of an electron 
respectively.
Here $c_{i}$ ($c_{i}^{\dag}$) are the fermion annihilation (creation) operators 
for destroying (creating) electron at site '$i$' and similarly $d_{\alpha k}$ 
($d_{\alpha k}^{\dag}$) are operators for 
destroying (creating) electron in state 'k' in the '$\alpha$' lead 
($\alpha=L/R$). First two terms in the Hamiltonian represent free system and 
free lead Hamiltonians, and the third term represents hybridization 
between system and lead sites.
We also assumed wide-band approximation (i.e., system lead hybridization is 
independent of 'k'). 
\subsection*{Bond currents}
Expressions for bond current operators between localized sites can be obtained 
from the continuity equation for charge density operator at any localized site. 
For example 
rate of change of charge at site '1' i.e,
\begin{align}
\label{eq-3}
\frac{d }{dt}(-e c_1^{\dag}c_1^{})=&\frac{ie}{\hbar}[c_1^{\dag}c_1^{},H]
\end{align}
gives three terms on the R.H.S. and each of these three terms can be identified 
as the operator for current from site '1' to site '2' or to site '4' or to left 
lead regions. 
In particular, the operator for current from site '2' to '1' and from site '4' 
to '1'can be identified as,
\begin{align}
\label{eq-4}
 \hat I_{2\rightarrow1}=\frac{ie}{\hbar}\big( e^{i\frac{\phi}{4}}_{} 
c_2^{\dag}c_1^{}-e^{-i\frac{\phi}{4}}_{} c_1^{\dag}c_2^{} \big)
\end{align}
 and  
 \begin{align}
 \label{eq-5}
 \hat I_{4\rightarrow1}=\frac{ie}{\hbar}\big( e^{-i\frac{\phi}{4}}_{} 
c_4^{\dag}c_1^{}-e^{i\frac{\phi}{4}}_{} c_1^{\dag}c_4^{} \big)
\end{align}
whose average gives the bond currents flowing between sites '1' $\&$ '2' 
($I_{2\rightarrow1}$) and '1' $\&$ '4' ($I_{4\rightarrow1}$)). At steady state 
these two bond currents 
are expressed as,
\begin{eqnarray}
\label{eq-6}
 I_{2\rightarrow1}&=& 
\frac{e}{\hbar}\int_{-\infty}^{+\infty}\frac{d\omega}{2\pi}
\big[e^{i\frac{\phi}{4}}G^{<}_{12}(\omega)-e^{-i\frac{\phi}{4}}G^{<}_{21}
(\omega)\big]
\end{eqnarray}
and
\begin{eqnarray}
\label{eq-7}
 I_{4\rightarrow1}&=& 
\frac{e}{\hbar}\int_{-\infty}^{+\infty}\frac{d\omega}{2\pi}
\big[e^{-i\frac{\phi}{4}}G^{<}_{14}(\omega)-e^{i\frac{\phi}{4}}G^{<}_{41}
(\omega)\big]
\end{eqnarray}
where $G^{<}_{ab}$ is the 'ab' matrix element of Fourier transformed lesser 
projections of systems Green's function to be introduced shortly. 
 Note that these are the only independent bond currents flowing inside the 
molecule, as all other bond currents can be expressed in terms of these two 
currents due to stationarity 
of charge densities at all the sites in the molecule at steady-state. Indeed, at steady state  
$I_{2 \rightarrow 5}=0$, $I_{3 \rightarrow 2}=I_{2 \rightarrow 1}$ and 
$I_{3 \rightarrow 4}=I_{4 \rightarrow 1}$.
\subsection*{Net current in the circuit}
The net current $I_L$ (which is same as $-I_R$) flowing into the left lead from 
site '1' at steady-state is given by the rate of change of charge on the left 
lead i.e.,
$I_L(t)=\frac{d}{dt}(-e\sum_k d_{LK}^\dag d_{LK})$. Similar to the bond 
currents, the net current can also be expressed in terms of system greater and 
lesser Green's functions  
$G^{>/<}$ as \cite{haug},
\begin{align}
\label{eq-8}
I_L=\frac{e}{\hbar}\int_{-\infty}^{+\infty}\frac{d\omega}{2\pi}\big[\Sigma^{<}_{
11}(\omega)
G^{>}_{11}(\omega)-G^{<}_{11}(\omega)\Sigma^{>}_{11}(\omega)\big], 
\end{align}
where $\Sigma^{>/<}$ are Fourier transformed greater and lesser projections of 
contour ordered self energy to be introduced shortly. 
Note that the two terms on the R.H.S. of Eq.(\ref{eq-8}) are real and represent 
inflow 
and out flow of the electrons from the left lead. On the other hand, such an
interpretation is not possible for the two terms on the R.H.S. of 
Eq.(\ref{eq-6})
and Eq.(\ref{eq-7}) as they are complex functions in general.
\subsection*{Green's function calculation}
In order to calculate bond currents and net current in the circuit, we need to 
compute the system's Green's functions. These Green's functions (in matrix 
form) 
are defined on Schwinger-Keldysh contour \cite{rammer,haug} as,
\begin{eqnarray}
\label{eq-9}
 &&G^{c}(\tau,\tau')=\nonumber\\
 &&-\frac{i}{\hbar}\langle\big[\Theta(\tau,
\tau')\Psi(\tau)\Psi^\dag(\tau')-\Theta(\tau',
\tau)\Psi^\dag(\tau')^T\Psi(\tau)^T\big]\rangle
\end{eqnarray}
where $\tau$ and $\tau'$ are contour times with,
\begin{eqnarray}
\label{eq-10}
 \Psi(\tau)&=&\begin{pmatrix}
             c_1^{}(\tau) & c_2^{}(\tau) & c_3^{}(\tau) & c_4^{}(\tau) & 
c_5^{}(\tau)\\
            \end{pmatrix}^T
\end{eqnarray}
and $\Theta(\tau,\tau')$ is the Heaviside step function defined on the
Schwinger-Keldysh contour \cite{rammer}.
$G^{c}(\tau,\tau')$ satisfies the following equation of motion 
\cite{rammer,haug}
\begin{eqnarray}
\label{eq-11}
 &&\int_c 
d\tau_1\big[(i\hbar\frac{\partial}{\partial\tau}-H_{S})\delta^{c}(\tau,
\tau_1)-\Sigma_{}^{c}(\tau,\tau_1)\big]G^{c}(\tau_1,\tau')=\delta^{c}(\tau,
\tau')
\end{eqnarray}
where $\Sigma^{c}$ is the self-energy due to interaction with the leads and has 
the
following matrix structure
{\scriptsize
\begin{eqnarray}
\label{eq-12}
&&\Sigma^{c}_{}(\tau,\tau')=\nonumber\\
&&\begin{bmatrix}                  
|g_{L}^{}|^2\displaystyle\sum_{k,k'}^{}G^0_{Lk,Lk'}(\tau,\tau')&0&0&0&0\\
0&0&0&0&0\\                   
0&0&|g_{R}^{}|^2\displaystyle\sum_{k,k'}^{}G^0_{Rk,Rk'}(\tau,\tau')&0&0\\
0&0&0&0&0\\
0&0&0&0&0\\
\end{bmatrix}.\nonumber\\
\end{eqnarray} 
}
Here $G^0_{Lk,Lk'}(\tau,\tau')$ and $G^0_{Rk,Rk'}(\tau,\tau')$ are contour
ordered Green's functions for the isolated leads. Equation (\ref{eq-11}) can be 
projected onto 
the real times using Langreth rules to obtain all other real-time Green's
functions\cite{haug}. At steady-state all the Green's functions become time 
translational
invariant and can be handled easily in the frequency domain. For example, 
equation for retarded system Green's function can be obtained from 
Eq. (\ref{eq-11}) by using Langreth rules and Fourier transforming the 
resulting 
equation to get,
\begin{eqnarray}
\label{eq-13}
 \big[\omega I -H_0 -\Sigma^r(\omega)\big]G^r_{}(\omega)=I,
\end{eqnarray}
where '$I$' is a $5\times5$ identity matrix and $\Sigma^r(\omega)$ is Fourier 
transformed retarded self energy, obtained by Fourier transforming retarded 
projection of contour ordered self 
energy $\Sigma^{c}_{}(\tau,\tau')$ given in Eq. (\ref{eq-12}). 
Retarded Green's function can be obtained from the above equation by matrix 
inversion i.e., $G^r_{}(\omega)=\big[\omega I -H_0 
-\Sigma^r(\omega)\big]^{-1}$. 
Advanced Green's function can be obtained in a similar manner, i.e., 
$G^a_{}(\omega)=\big[\omega I -H_0 -\Sigma^a(\omega)\big]^{-1}$, where 
$\Sigma^a(\omega)$ is Fourier transformed advanced self energy. Lesser and 
greater Green's functions can be 
obtained from, 
\begin{eqnarray}
\label{eq-14}
 G^{</>}_{}(\omega)&=&G^{r}_{}(\omega)\Sigma^{</>}_{}(\omega)G^{a}_{}(\omega)
\end{eqnarray}
where $\Sigma^{</>}_{}(\omega)$ are Fourier transformed lesser and greater self 
energies obtained by Fourier transforming lesser and greater projections of 
contour ordered self energy given in Eq. (\ref{eq-12}).\par
\indent
Thus obtained Green's functions can be used in Eqs. (\ref{eq-6}), (\ref{eq-7}) 
and (\ref{eq-8}) to get expressions for the bond currents, $I_{2\rightarrow1}$ 
and $I_{4\rightarrow1}$ and the net current, $I_{L}$, as (from here onwards we 
choose units such that $e=1$ and $\hbar=1$),
\begin{widetext}
\begin{eqnarray}
\label{eq-15}
I_{2\rightarrow1}&=&\int_{-\infty}^{+\infty}\frac{d\omega}{2\pi}\frac{
\Gamma_L\Gamma_R[f_L(\omega)-f_R(\omega)]}{D[\omega]}\omega(\omega-\epsilon)\big
[\{2\omega(\omega-\epsilon)-t^2\}\cos^2(\frac{\phi}{2})+t^2\sin^2(\frac{\phi}{2}
)\big]\nonumber\\
&+&\int_{-\infty}^{+\infty}\frac{d\omega}{2\pi}\frac{2[
\Gamma_Lf_L(\omega)+\Gamma_Rf_R(\omega)]\sin(\phi)}{D[\omega]}
(\omega-\epsilon)\big[\omega(\omega-\epsilon)(\omega^2-2)-t^2(\omega^2-1)\big],
\end{eqnarray}
\begin{eqnarray}
\label{eq-16}
I_{4\rightarrow1}&=&\int_{-\infty}^{+\infty}\frac{d\omega}{2\pi}\frac{
\Gamma_L\Gamma_R[f_L(\omega)-f_R(\omega)]}{D[\omega]}\{
\omega(\omega-\epsilon)-t^2\}\big[\{2\omega(\omega-\epsilon)-t^2\}\cos^2(\frac{
\phi}{2})-t^2\sin^2(\frac{\phi}{2})\big]\nonumber\\
&-&\int_{-\infty}^{+\infty}\frac{d\omega}{2\pi}\frac{2[
\Gamma_Lf_L(\omega)+\Gamma_Rf_R(\omega)]\sin(\phi)}{D[\omega]}
(\omega-\epsilon)\big[\omega(\omega-\epsilon)(\omega^2-2)-t^2(\omega^2-1)\big],
\end{eqnarray}
and 
\begin{eqnarray}
\label{eq-17}
I_{L}&=&\int_{-\infty}^{+\infty}\frac{d\omega}{2\pi}\frac{\Gamma_L\Gamma_R[
f_L(\omega)-f_R(\omega)]}{D[\omega]}\big[\{2\omega(\omega-\epsilon)-t^2\}
^2\cos^2(\frac{\phi}{2})+t^4\sin^2(\frac{\phi}{2})\big].
\end{eqnarray}
\end{widetext}
Here 
$D[\omega]=\big[(\omega-\epsilon)\{\omega^4-(\frac{\Gamma_L\Gamma_R}{4}
+4)\omega^2+4\sin^2(\frac{\phi}{2})\}-\omega 
t^2\{\omega^2-2-\frac{\Gamma_L\Gamma_R}{4}\}\big]^2+(\frac{\Gamma_L+\Gamma_R}{2}
)^2\big[\omega(\omega-\epsilon)(\omega^2-2)-t^2(\omega^2-1)\big]^2$, 
$\Gamma_{\alpha}=2\pi\rho|g_{\alpha}|^2$ 
and $f_{\alpha}(\omega)=\frac{1}{e^{\beta_{\alpha}(\omega-\mu_{\alpha})}+1}$. 
Note that $\Gamma$, $t$ and $\omega$ are dimensionless numbers given in units 
of 
the coupling between sites constituting the ring. Both the bond 
currents, $I_{2 \rightarrow 1}$ and $I_{4 \rightarrow 1}$ have two 
contributions, one purely 
due to applied bias (and becomes zero for $eV=0$) and the other purely due to 
applied magnetic 
flux (and becomes zero for $\phi=0$). 
In passing, we note that $I_L= I_{2\rightarrow1} + I_{4\rightarrow1}$, which is 
nothing 
but Kirchoff's law. Notice, the net transmission function given as,
\begin{eqnarray}
\label{eq-17+1}
&&T_{L}(\omega)=\frac{\Gamma_L\Gamma_R}{D[\omega]}\big[\{
2\omega(\omega-\epsilon)-t^2\}^2\cos^2(\frac{\phi}{2})+t^4\sin^2(\frac{\phi}{2}
)\big],
\end{eqnarray}
has no real zeros (anti resonances) for $\phi\neq 2n\pi$ ('n' is any integer). 
For $\phi= 2n\pi$, $T_{L}(\omega)$ has zeros at 
$\omega=\frac{\epsilon\pm\sqrt{\epsilon^2+2 t^2}}{2}$.\par
\indent
The two different cases, symmetric and asymmetric junctions, mentioned in the 
introduction, are special cases of the model presented in this section. They 
are 
obtained in the limits $t\to0$ and $\phi\to0$, respectively. We analyze these two 
cases separately in the next two sections.
\section{Symmetric Aharonov-Bohm ring}
\label{sym-abring}
In this section we analyze the effect of applied magnetic field and bias 
on the bond currents flowing in a symmetric ring. We therefore take the limit 
of $t\to0$ in the general equations (\ref{eq-15}), (\ref{eq-16}) and 
(\ref{eq-17}), given in Section II. The extra site (substituent) gets 
decoupled from the ring and hence does not affect the bond currents as well as 
the net current. The quantum Aharonov-Bohm effects on the net conductance of 
this 
junction is studied in Ref.\cite{zeng}, where the effect of magnetic flux and 
asymmetry between two branches on
the net transmission function were analyzed. In this work we are mainly 
interested in controlling bond currents inside the molecule. For simplification 
we set $\Gamma_L=\Gamma_R=\Gamma$.\par
\indent
For this symmetric Aharonov-Bohm ring case, bond currents become 
$I_{2\rightarrow1}=I_{V}+I_{\phi}$ and $I_{4\rightarrow1}=I_{V}-I_{\phi}$,
where $I_{V}$ and $I_{\phi}$ are given by 
\begin{eqnarray}
\label{eq-18}
&&I_{V}=\int_{-\infty}^{+\infty}\frac{d\omega}{2\pi}\bigg[\frac{
2\Gamma^2\omega^2\cos^2(\frac{\phi}{2})}{D[\omega]}\bigg]\big[
f_L(\omega)-f_R(\omega)\big]
\end{eqnarray}
and
\begin{eqnarray}
\label{eq-19}
 &&I_{\phi}=
\int_{-\infty}^{+\infty}\frac{d\omega}{2\pi}\bigg[\frac{
2\Gamma\omega(\omega^2-2)\sin(\phi)}{D[\omega]}\bigg]\big[
f_L(\omega)+f_R(\omega)\big]
\end{eqnarray}
with
$D[\omega]=\big[\omega^4-(\frac{\Gamma^2}{4}+4)\omega^2+4\sin^2(\frac{\phi}{2}
)\big]
^2+\Gamma^2\omega^2\big[\omega^2-2\big]^2$. 
%, $\Gamma=2\pi\rho|g|^2$ 
The expressions for $I_{2\rightarrow1}$ and $I_{4\rightarrow1}$ have two 
contributions : $I_{V}$ is
due to the applied chemical potential difference between two metallic leads and
$I_{\phi}$ is the contribution 
driven due to the magnetic flux, this contribution vanishes only if $\phi$ is an
integral multiple of $2\pi$. Note that both the contributions vanish if $\phi$
is an odd integral multiple of $\pi$, irrespective 
of the applied bias. This behavior can be understood better if we analyze the 
bond current in the molecular eigenspace (Appendix A) and the net current in terms of  
spatial pathways (Appendix B). We find that the two 
contributions, 
$I_{\phi}$ and $I_{V}$, have different origins. Each eigenstate carries a 
current 
which depends on $\phi$. These add up to give $I_{\phi}$, while $I_{V}$ 
contribution 
comes due to the coherences induced by the leads between different eigenstates. 
$I_{V}$ can also be interpreted as the net current 
due to two interfering pathways $1\to 2\to 3$ and $1\to 4\to 3$ in the molecule 
(Appendix B). 
At $\phi=\pi$, these two pathways interfere destructively and hence $I_{V}=0$. On the
other hand, $I_{\phi}=0$ for $\phi=\pi$, as eigenstates which carry opposite currents 
become degenerate (Appendix A). When $\phi=0$, the bond current, $I_{12}=I_{V}$, i.e., the 
contribution comes entirely from the coherences between eigenstates which can 
not be
described within the simplified Lindblad Quantum Master Equation approach \cite{Archak}. Analytical 
expressions for $I_{\phi}$ and $I_{V}$ for both finite temperature and zero 
temperature 
cases are given in Appendix C. Note that for the bias driven part, $I_{V}$, it 
is straightforward 
to define an energy dependent transmission function $T(\omega)=\bigg[\frac{
2\Gamma^2\omega^2\cos^2(\frac{\phi}{2})}{D[\omega]}\bigg]$, however the same is 
not 
possible for $I_{\phi}$.\par
\indent Close to equilibrium, by linearizing the two fluxes in 
$\phi$ and $eV$, we get : 
\begin{eqnarray}
\label{eq-20}
 I_{V}=L_{VV}\times eV+ L_{V\phi}\times\phi\\
 I_{\phi}=L_{\phi V}\times eV+ L_{\phi\phi}\times\phi
 \end{eqnarray}
 where $L_{VV}=\int_{-\infty}^{+\infty}\frac{d\omega}{2\pi}\bigg[\frac{
2\Gamma^2\omega^2}{D[\omega]_{\phi=0}}\bigg]f'(\omega)$, 
$L_{\phi\phi}=\int_{-\infty}^{+\infty}\frac{d\omega}{2\pi}\bigg[\frac{
4\Gamma\omega(\omega^2-2)}{D[\omega]_{\phi=0}}\bigg]f(\omega)$ and 
$L_{V\phi}=L_{\phi V}=0$ are 
Onsager matrix elements. The off-diagonal elements are individually zero since, 
(i) $I_{V}$ is an even function of $\phi$ and $I_{V}=0$ for $eV=0$, hence 
contribution linear in $\phi$ to $I_{V}$ vanishes, 
and (ii) $I_{\phi}$ is an even function of $eV$ (for $\mu=0$) and $I_{\phi}=0$  
for $\phi$, hence contribution linear in $eV$ to $I_{\phi}$ vanishes. Thus 
close 
to equilibrium the two fluxes, one originating from the applied bias and the 
other due to the 
applied magnetic field, are independent of each other. Therefore, close to 
equilibrium, the net current in the circuit 
($2I_{V}$) can not be manipulated by applied magnetic field. Note that here, 
$\phi$ acts as a thermodynamic force for the flux $I_{\phi}$. This scenario 
is different from standard linear irreversible thermodynamics, where the cross 
Onsager matrix elements for a general case, where generalized fluxes $J_m$, are 
driven by generalized forces $X_n$ (i.e., $J_m=\sum_n L_{mn} X_n$), satisfy 
Onsager-Casimir relationship \cite{Groot} as a consequence of microscopic 
reversibility of underlying dynamics, 
$L_{mn}(\phi)= (-1)^{(\alpha_m+\alpha_n)}L_{nm}(-\phi)$ ($\alpha_m$ assumes '0' 
if  $J_m$ and $X_m$ are symmetric under time reversal or '1' if $J_m$ and $X_m$ 
are antisymmetric under time reversal),  where $\phi$ is treated as a 
parameter. 
Since here $\phi$ is an 
external force which drives $I_{\phi}$, on time reversal both the force 
($\phi$) 
and hence the resultant flux ($I_{\phi}$) change sign, which is consistent with 
linear irreversible thermodynamics \cite{Groot}. 
\subsection*{Thermodynamic equilibrium}
\label{equilibrium}
When $\mu_L=\mu_R=\mu$ and $\beta_L=\beta_R=\beta$, (i.e., when both the leads 
are at the same thermodynamic 
equilibrium) only the magnetic field driven current
$I_{\phi}=I_{2\rightarrow1}$ exists ($I_{4\rightarrow1}=-I_{2\rightarrow1}$) 
and leads only act as phase-breakers for the electronic motion in the molecular 
ring. Note that in this case $\phi$ may be arbitrarily large. \par
\indent
\begin{figure}[!htbp]
\centering
\includegraphics[width=7.5cm,height=5.5cm]{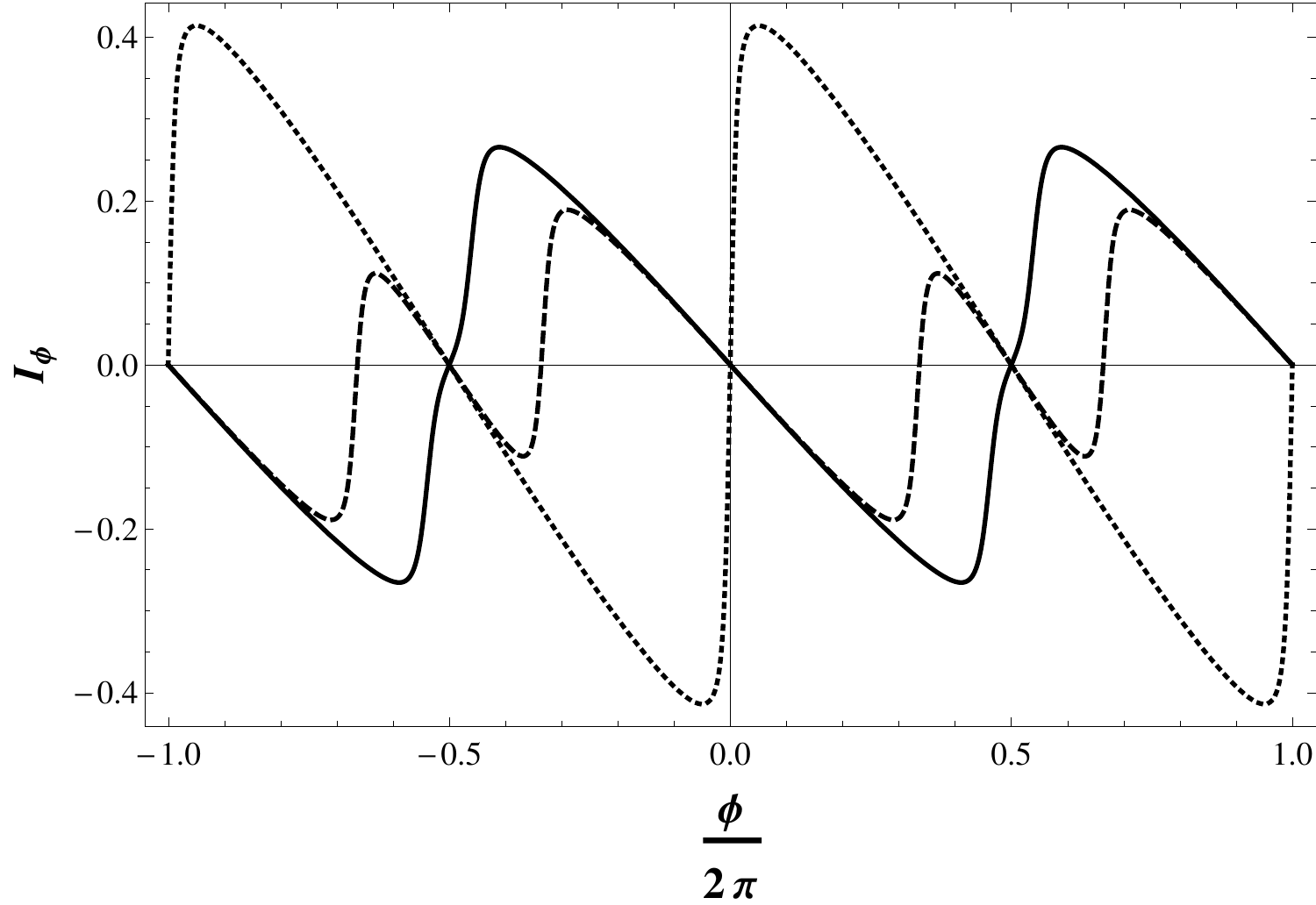}
\caption{Equilibrium bond current as a function of $\phi$ with $\Gamma=0.1$ and
$\beta=100$. Here, dashed : $\mu=-1.0$, dotted : $\mu=0$ and continuous : 
$\mu=1.5$. Note that all energy values are given in units of coupling strength 
between sites constituting the ring.}
\label{fig-3}
\end{figure}
Figure \ref{fig-3} is a plot of the 
equilibrium bond current, $I_{\phi}$, as a function of $\phi$ for various 
values of 
the chemical potential ($\mu$) of leads at fixed $\Gamma$ and $\beta$. It shows
that at thermodynamic equilibrium, $I_{\phi}$ is a periodic function
of $\phi$ with period $2\pi$, although eigenstate energies, eigenstate contributions 
(to $I_{phi}$) and their populations, are periodic in $\phi$ with period $8\pi$. This 
is because the eigenstate populations and their respective 
contributions to $I_{\phi}$ get swapped after $2\pi$ increase in $\phi$ such 
that $I_{\phi}$ remains periodic in $\phi$ with period $2\pi$ (appendix A).\par
\indent
\begin{figure}[!htbp]
\centering
\includegraphics[width=7.5cm,height=5.5cm]{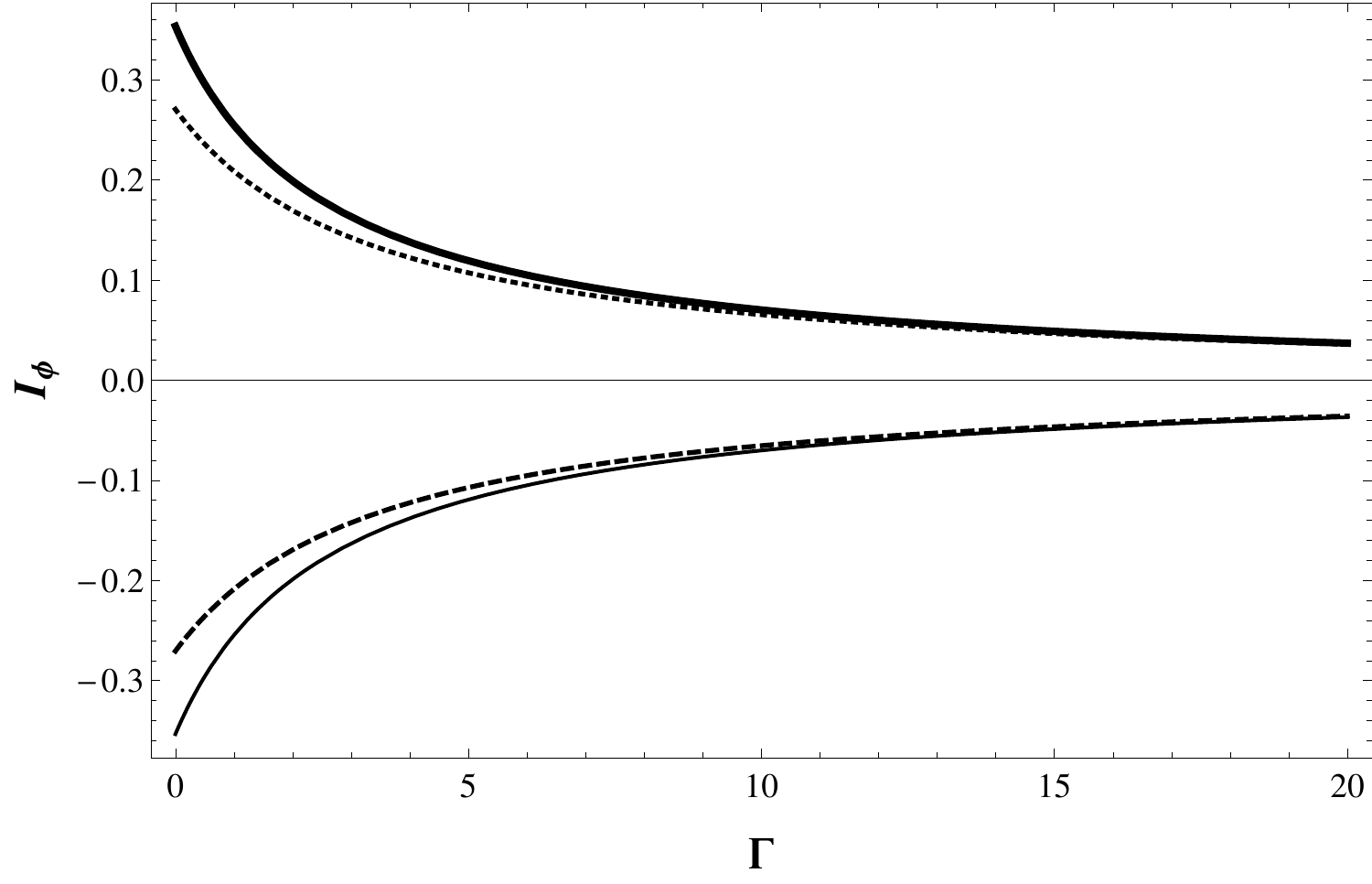}
\caption{Equilibrium bond current as a function of $\Gamma$ with bare chemical
potential $\mu=0$ and $\beta=100$. Here, dashed : $\phi=-\frac{\pi}{2}$, thin 
: $\phi=-\frac{\pi}{3}$, thick : $\phi=\frac{\pi}{3}$ and dotted : 
$\phi=\frac{\pi}{2}$.}
\label{fig-4}
\end{figure}
We next analyze the effect of molecule-lead coupling on the equilibrium bond 
current. 
Figure \ref{fig-4} shows $I_{\phi}$ as a
function of $\Gamma$ at fixed $\mu$ and $\beta$ for various 
values of $\phi$. Increasing the coupling strength to leads,
$I_{\phi}$ decreases because leads acts as phase breakers that hinder the 
coherent
motion of electrons \cite{Buttiker1,Buttiker2} and therefore suppresses the 
coherent current. 
As $\Gamma$ is increased, different eigenstates mix strongly due to 
coupling to leads. This enhances scattering of electrons 
between different eigenstates and leads to dephasing. Said differently, the 
suppression of $I_{\phi}$ can also be understood as due to increasing overlap 
between density of states of different eigenstates carrying opposite currents, 
as $\Gamma$ is increased. 
As $\Gamma \rightarrow \infty$ (specifically $\Gamma^2\gg\Gamma\beta\gg1$), $I_{\phi}$ decays to 
zero as,
\begin{eqnarray}
\label{eq-21}
I_{\phi}&\approx&\frac{4\beta 
\sin(\phi)}{\pi^2\Gamma^2}Re[\Psi^{(1)}[\frac{1}{2}-\frac{i\beta\mu}{2\pi}]],
\end{eqnarray}
where $\Psi^{(1)}[Z]$ is trigamma function \cite{Thomson} in variable '$Z$'. 
On the other hand, as $\Gamma\rightarrow 0$, $I_{\phi}$ reduces to the 
limit of circulating current in an isolated molecule which is given by the sum 
of the currents carried by eigenstates (Appendix A) multiplied by their 
respective populations (at thermodynamic equilibrium given by lead Fermi functions 
at the corresponding eigenstate energies). 
\par
\indent
\begin{figure}[!htbp]
\centering
\includegraphics[width=7.5cm,height=5.5cm]{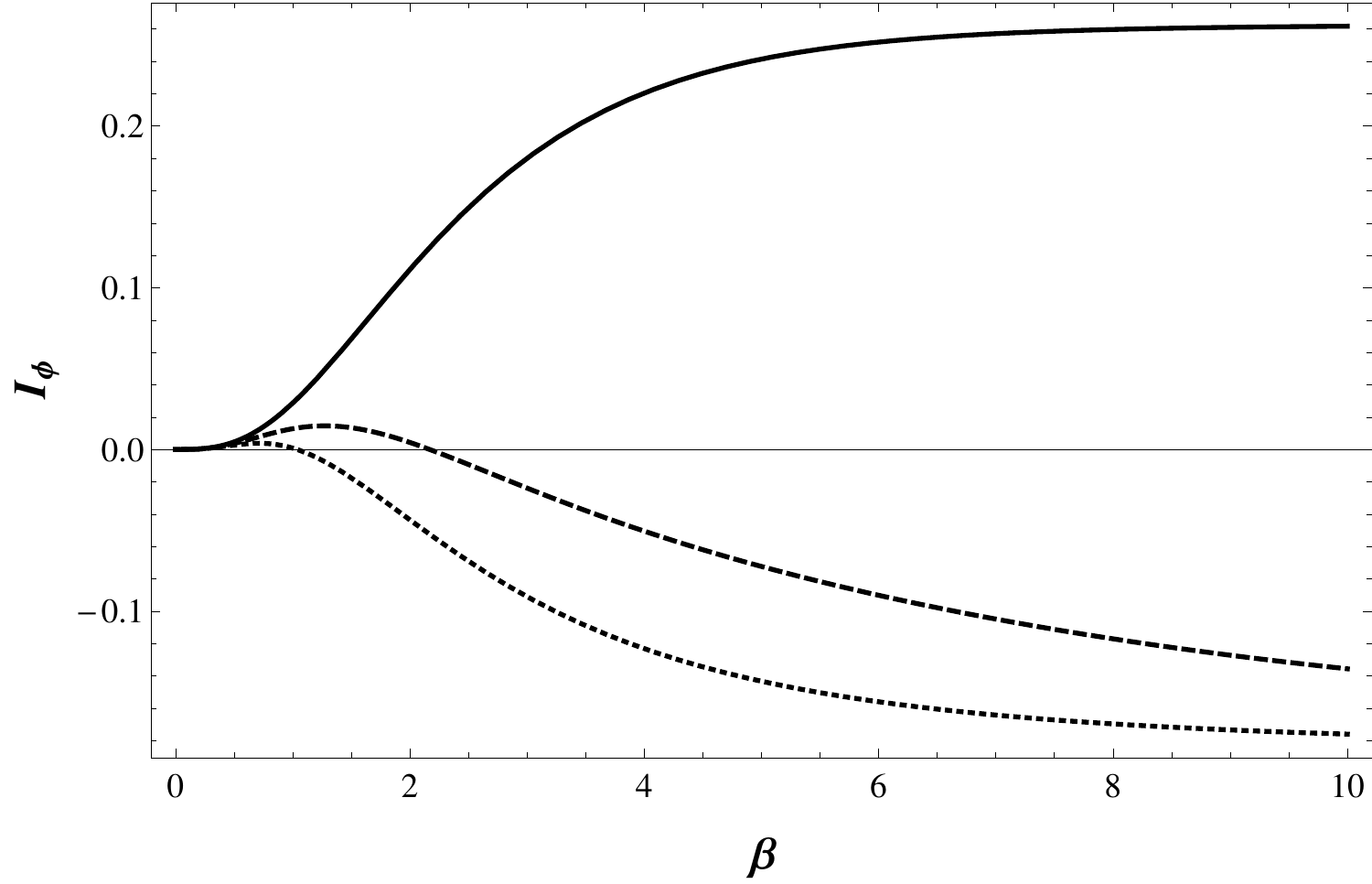}
\caption{Equilibrium bond current as a function of $\beta$ with $\Gamma=0.1$ and
$\phi=\frac{\pi}{2}$. Here, dashed : $\mu=-1.0$, continuous : $\mu=0$ and 
dotted 
: $\mu=1.5$.}
\label{fig-5}
\end{figure}
As the temperature is increased, different eigenstates start to get populated 
due to 
coupling with leads. At high temperatures ($\beta \rightarrow 0$), the 
populations of 
various eigenstates become almost identical. As discussed in Appendix A, 
different eigenstates contribute oppositely 
to the current, and hence the net bond current diminishes as the temperature is 
increased. 
This is shown in Fig. \ref{fig-5}. At small 
temperature ($\beta \rightarrow \infty$), the current approaches to the sum of 
currents carried by all eigenstates with 
energies below $\mu$ (since only states below $\mu$ are occupied).
\subsection*{Out of thermodynamic equilibrium}
We now consider the case when the two leads are not at the thermodynamic 
equilibrium, i.e., $\mu_L\neq\mu_R$ and $\beta_L=\beta_R=\beta$. 
Further, we take $\mu_L=\mu+\frac{eV}{2}$ and $\mu_R=\mu-\frac{eV}{2}$, with 
bare chemical potentials of the leads set in resonance with ring site energies ($\mu=0$). 
In this case, the external bias also contributes to the bond currents and we have 
to consider both $I_{\phi}$ and $I_{V}$ given in Eqs. (\ref{eq-18}) and 
(\ref{eq-19}). 
We note that $I_{V}$ is an even (odd) function of $\phi$ ($eV$), 
whereas $I_{\phi}$ is an odd (even) function of $\phi$ ($eV$). Hence, by 
changing the polarity of either $eV$ or $\phi$, it would be possible to make the
two contributing currents flow in opposite directions leading to enhancement of 
current flowing along one branch and reduction of current flowing along the 
other branch. This is a trivial case.  
\begin{figure}[!htbp]
\includegraphics[width=7.5cm,height=5.5cm]{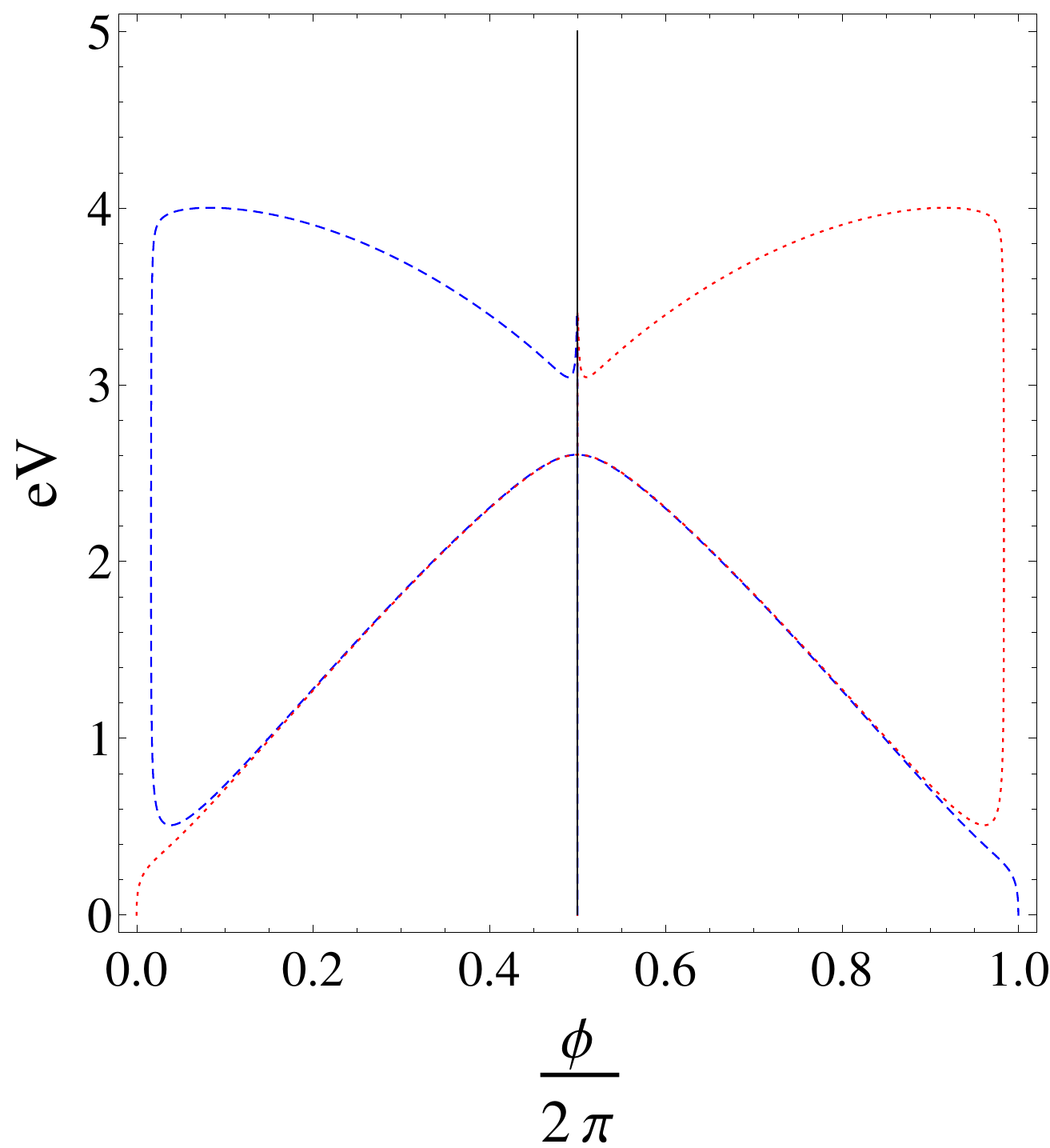}
\caption{(Color online) Phase diagram for bond currents in the molecular ring. 
The two curves (red-dotted and blue-dashed) separate regions 
where both the branches are conducting and black line represents a region where 
both the branches are non conducting. On the blue (dashed) curve only lower 
branch is conducting, while on the red (dotted) curve upper branch is 
conducting. 
Parameters chosen are $\Gamma=0.1$, $\beta=100$, $\mu=0$, $\mu_L=\mu+eV/2$ and 
$\mu_R=\mu-eV/2$.}
\label{fig-6e}
\end{figure}
However we find that, at finite bias ($eV$), $\phi$ can be tuned (without changing polarity) 
such that only one of the branches is conducting. 
This is shown in Fig.\ref{fig-6e}, where the white region in $\phi$-$eV$ space 
indicates that both the 
branches are conducting, the blue (dashed) curve corresponds to $\phi$ and $eV$ 
values where only lower branch is conducting ($I_{2\rightarrow 1}=0$) and the 
red (dotted) curve 
corresponds to $\phi$ and $eV$ values where only the upper branch is conducting 
($I_{4\rightarrow 1}=0$). It should be recalled that 
at $\phi=\pi$, due to destructive interference, both the branches are non conducting and 
hence net current in the circuit is also zero (Appendix B), irrespective 
of the applied bias. This is indicated by a black line in the figure. 
\begin{figure}[!htbp]
\includegraphics[width=8.0cm,height=14.0cm]
{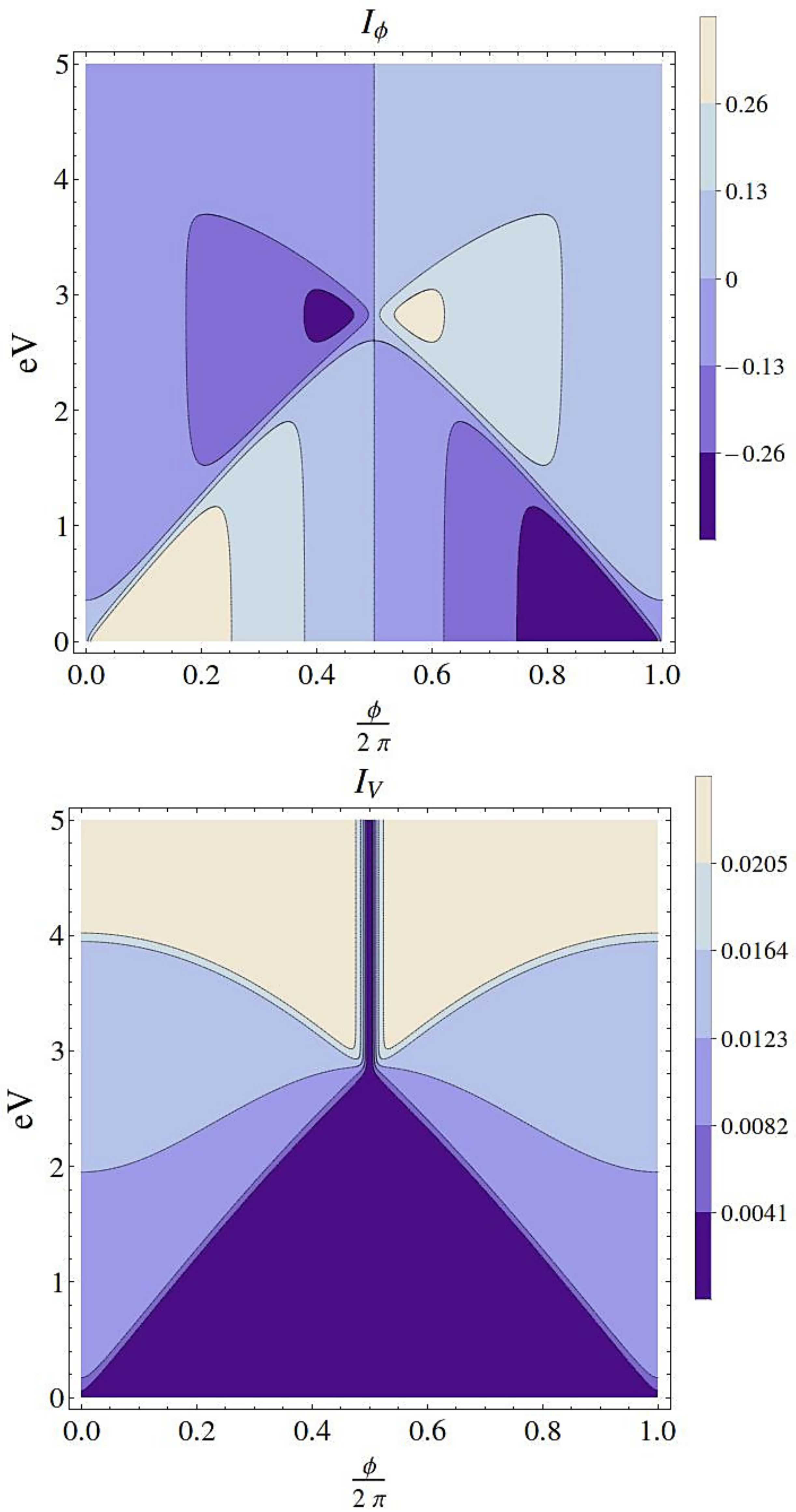}
\caption{(Color online) Contributions, $I_{\phi}$ (upper panel) and $I_{V}$ 
(lower panel), to bond currents are plotted as a function of 
$\phi$ and $eV$ with $\Gamma=0.1$, $\beta=100$, $\mu=0$, $\mu_L=\mu+eV/2$ and 
$\mu_R=\mu-eV/2$.}
\label{fig-6}
\end{figure}
Magnetic field driven ($I_{\phi}$) 
and applied bias driven ($I_{V}$) contributions to  the bond current are plotted in 
Fig.\ref{fig-6} as a function of $\phi$ and $eV$.  
Notice that $I_{\phi}$ changes sign with respect to both $\phi$ and $eV$, while 
direction of $I_{V}$ cannot be 
changed by changing $\phi$. The change in the sign of $I_{\phi}$ with $\phi$ is 
similar to the case discussed in the 
thermodynamic equilibrium. However the change in the sign of $I_{\phi}$ with 
$eV$ happens because, as the bias 
increases, the populations of states with different contributions changes, resulting in 
sign change of $I_{\phi}$.\par 
\indent The net current flowing in the circuit for the symmetric ($t=0$) Aharonov-Bohm 
ring case becomes,
\begin{eqnarray}
\label{eq-22}
I_L&=&\int_{-\infty}^{+\infty}\frac{d\omega}{2\pi}\bigg[\frac{
4\Gamma^2\omega^2\cos^2(\frac{\phi}{2})}{D[\omega]}\bigg][
f_L(\omega)-f_R(\omega)].
\end{eqnarray}
Net current flowing in the circuit has been analyzed in 
several works to study the effects of the magnetic flux on the net current, for
example, Refs. \cite{rai,zeng} studied the effect of magnetic field on net 
transmission function in asymmetric ring system. 
In Ref. \cite{Bai} the effect of inhomogeneous magnetic flux on the net current 
is analyzed. 
In Ref. \cite{Sztenkiel} the effect of coulomb interaction on the net current 
in 
the presence of 
magnetic flux is studied. Dissipation due to electron-phonon coupling and its 
effect on the net transmission 
has been studied in Ref. \cite{Wohlman2}, and the effect of
external electromagnetic field has been discussed in Ref. \cite{Wohlman1}. 
In the present work, since we are only interested in bond currents in the 
molecule, we do not pursue the net current further. 
\section{Effect of chemical substitution}
\label{chem-subs}
Here we explore the effect of coupling an extra
site (with site energy '$\epsilon$' and coupling strength 't') to an otherwise 
symmetric ring system. We 
analyze the effect of this substitution on the bond currents in the absence of 
magnetic field. 
The substitution introduces asymmetry between two paths that an electron 
can take in going from the left lead to the right lead. This leads to 
interference effects 
in the net current as discussed in Refs.\cite{Jayannavar2,Jayannavar3}. However 
this asymmetry not 
only affects the net current but also the bond currents in the molecule and may 
lead to circulating 
currents (at finite bias) even in the absence of magnetic flux. The goal in this section is 
to study these circulating currents. To this end we take the limit $\phi\to0$ of the general 
equations (\ref{eq-15}), (\ref{eq-16}) and (\ref{eq-17}), given in Section. II. 
To further simplify the analysis, we consider the case 
$\Gamma_L=\Gamma_R=\Gamma$ and $\epsilon=0$.\par
\begin{center}
\begin{figure}[!htbp]
\includegraphics[width=8.0cm,height=19.5cm]
{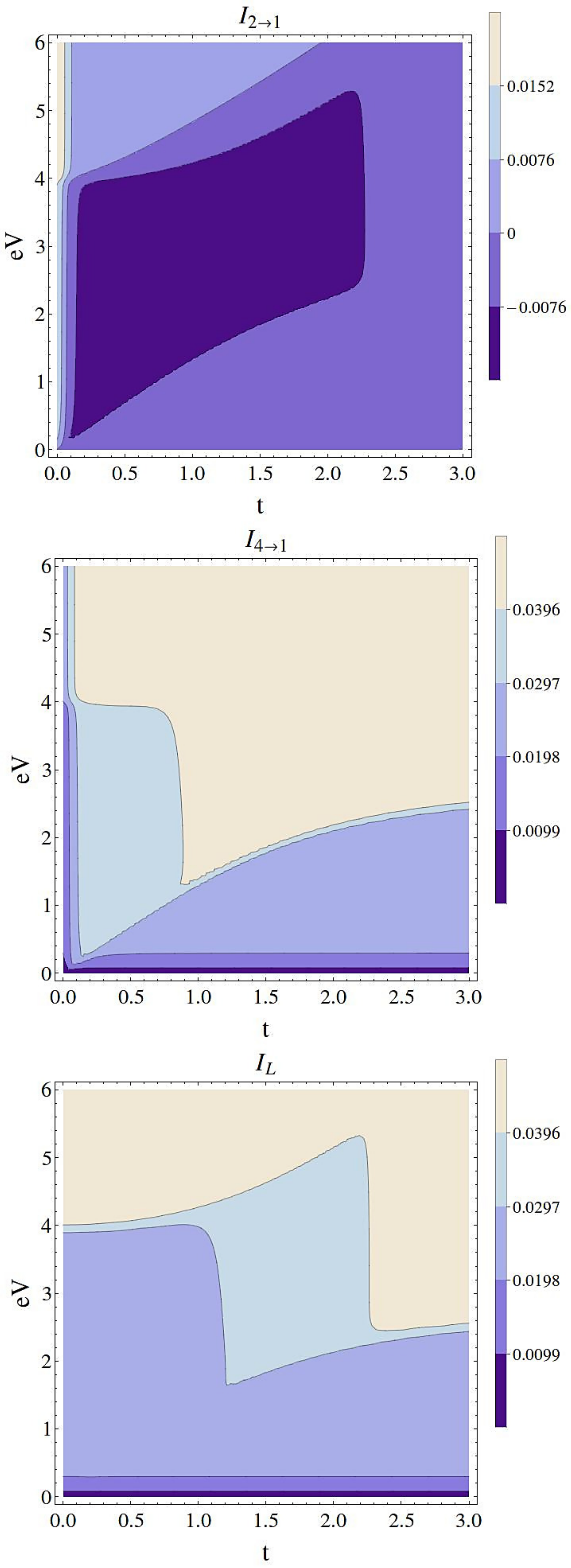}
\caption{(Color online) Bond currents $I_{2\rightarrow1}$ and 
$I_{4\rightarrow1}$ together with the net current, $I_L$, as a function of $t$ 
and $eV$ with $\epsilon=0$, $\Gamma=0.1$,
$\beta=100$, $\mu=0$, $\mu_L=\mu+eV/2$ and $\mu_R=\mu-eV/2$.}
\label{fig-8}
\end{figure}
\end{center}

\indent
The expressions for the bond currents, 
$I_{2\rightarrow1}$ and $I_{4\rightarrow1}$ and the net current, $I_L$, assume 
the form,
\begin{eqnarray}
\label{eq-23}
&&I_{2\rightarrow1}=\int_{-\infty}^{+\infty}\frac{d\omega}{2\pi}\bigg[\frac{
\Gamma ^2 \omega ^2 (2 \omega ^2 - 
t^2)}{D[\omega]}\bigg]\big[f_L(\omega)-f_R(\omega)\big],\nonumber\\
\end{eqnarray}
\begin{eqnarray}
\label{eq-24}
&&I_{4\rightarrow1}=\int_{-\infty}^{+\infty}\frac{d\omega}{2\pi}\bigg[\frac{
\Gamma ^2 (\omega ^2 -t^2) (2 \omega ^2 - 
t^2)}{D[\omega]}\bigg]\big[f_L(\omega)-f_R(\omega)\big],\nonumber\\
\end{eqnarray}
\begin{eqnarray}
\label{eq-25}
&&I_L=\int_{-\infty}^{+\infty}\frac{d\omega}{2\pi}\bigg[\frac{\Gamma ^2 (2 
\omega ^2 - t^2)^2}{D[\omega]}\bigg]\big[f_L(\omega)-f_R(\omega)\big]
\end{eqnarray}
where 
$D[\omega]=\big[\omega^2+(\frac{\Gamma}{2})^2\big]\big[
((\omega^2-t^2)(\omega^2-4)-2 
t^2)^2+(\frac{\Gamma}{2})^2\omega^2(\omega^2-t^2)^2\big]$. 
These currents are plotted as functions of $eV$ and $t$ in Fig.\ref{fig-8}. 
Note that the bond current $I_{2\rightarrow1}$ (corresponding 
to the branch having the extra substituent) changes sign as bias is scanned, 
while $I_{4\rightarrow1}$ remains positive (in the direction of the net 
current).\par
\begin{figure}[!htbp]
\includegraphics[width=7.5cm,height=5.5cm]
{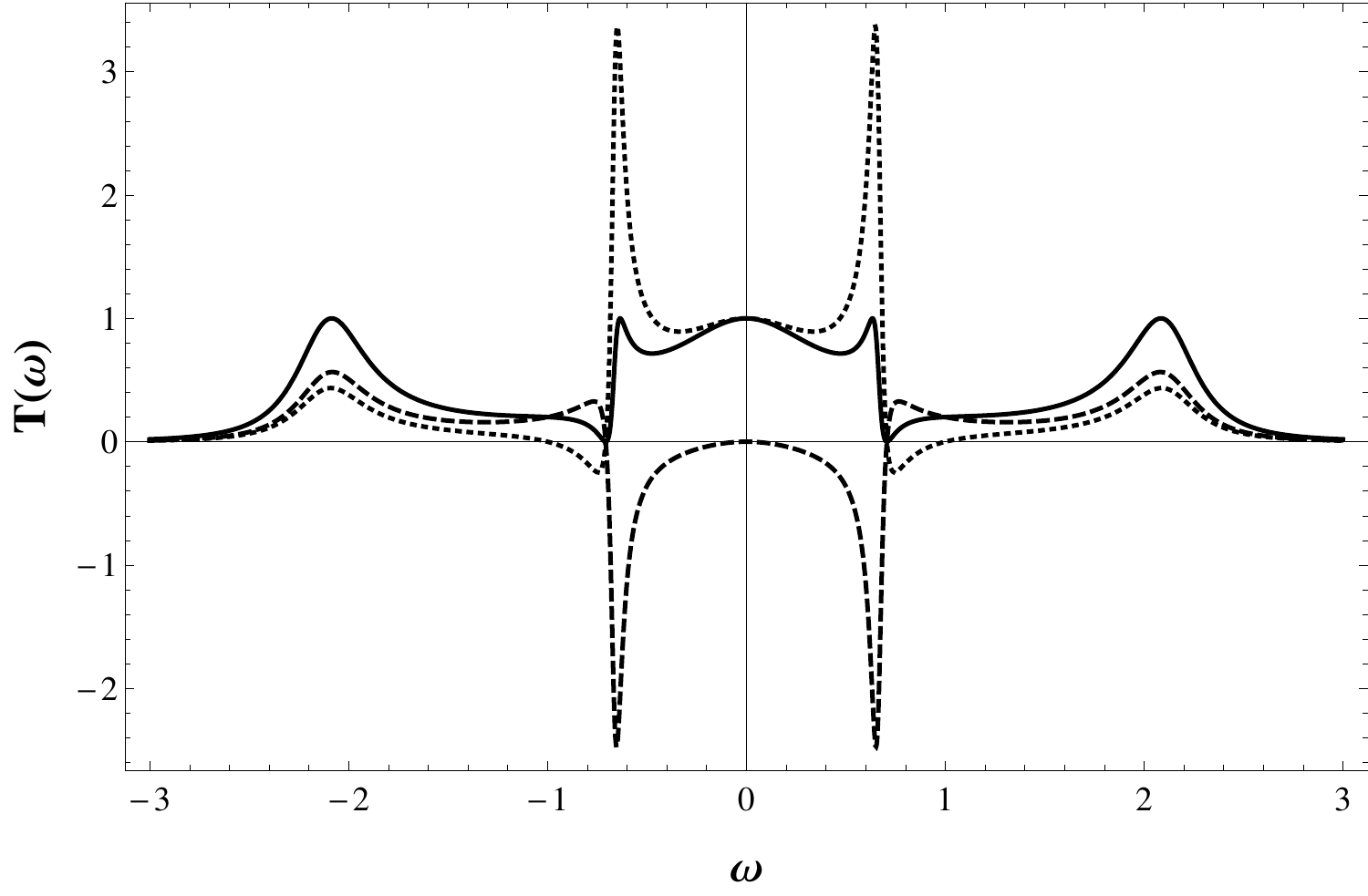}
\caption{Bond transmission functions $T_{12}$ (dashed), $T_{14}$ (dotted) and 
net transmission function $T_L$ (continuous) as a
function of $\omega$ with $\Gamma=1$, $\epsilon=0$ and $t=1$.}
\label{fig-9}
\end{figure}

\indent Unlike the case in presence of magnetic flux, in this case the two bond 
currents (which vanish at zero bias) have well defined energy dependent transmission functions, 
$T_{12}$ and $T_{14}$. We note that both the transmission functions have common 
zeros at $\omega=\pm\frac{t}{\sqrt{2}}$. We analyze 
the nature of transmission functions 
at these zeros. Since $D[\omega=\pm\frac{t}{\sqrt{2}}]>0$, it is clear that 
both 
$T_{12}$ and $T_{14}$ change sign in opposite directions 
around $\omega=\pm\frac{t}{\sqrt{2}}$. However the total transmission function, 
$T_L(\omega)=\frac{(2\omega^2-t^2)^2}{D[\omega]}$ attains 
its minimum value (zero) at these points (anti-resonances). This behavior 
of bond transmission functions changing sign around anti-resonances of the 
total transmission function was noticed 
by Jayannavar et.al., \cite{Jayannavar1} using scattering theory. Apart from 
the anti-resonance points, $T_{14}$ has extra 
zeros at $\omega=\pm t$, where it is an increasing (decreasing) function at 
$+t$ 
($-t$), and $T_{12}$ has an extra zero at $\omega=0$ where it has a maximum. 
For $|\omega|>t$, both $T_{12}(\omega)$ and $T_{14}(\omega)$ are positive 
functions of $\omega$. Thus at 
energies $|\omega|>t$, the two bond currents flow in the same (positive) 
direction, while for $|\omega|<t$, the two currents flow in the opposite 
direction and a circulating current exists. The energy range $|\omega|<t$ is, 
therefore, critical for the existence of a circulating current in the 
molecule. Thus at low temperatures, the circulating 
current exists only for $|eV|<t$ (here $\mu=0$ is assumed). In Fig.\ref{fig-9} 
we show a plot of 
$T_{12}(\omega)$, $T_{14}(\omega)$ and $T_{L}(\omega)$.\par 
\indent Zeros of transmission 
functions $T_{12}(\omega)$, $T_{14}(\omega)$ and $T_{L}(\omega)$ are analyzed 
using projection operator method in Appendix D. It is shown in Appendix D that the anti-
resonance (multi-path zero) of $T_{L}(\omega)$ at $\omega=\pm\frac{t}{\sqrt{2}}$ is due to the 
destructive interference between two paths '$1 \rightarrow 2 \rightarrow 3$' and 
'$1 \rightarrow 4 \rightarrow 3$' that an electron can take through the molecule to go from 
left lead to right lead. $T_{12}(\omega)$ and $T_{14}(\omega)$ are also zero at $\omega=\pm\frac{t}{\sqrt{2}}$,
but they do not have simple interpretation in this scheme. Zero (multi-path zero) of $T_{12}(\omega)$ at 
$\omega=0$ is (for $\epsilon=0$, another zero at $\omega=\epsilon$ called resonance zero coincides with zero at $\omega=0$), 
is due to the destructive interference between direct ('$1 \rightarrow 2$') and indirect ('$1 \rightarrow 4 \rightarrow 3 \rightarrow 2$') 
paths an electron can take to go from site '1' to site '2'. Zeros of $T_{14}(\omega)$ at $\omega=\pm t$ are due to the 
destructive interference between direct ('$1 \rightarrow 4$') and indirect paths ('$1 \rightarrow 2 \rightarrow 3 \rightarrow 4$') an electron 
can take to go from site '1' to site '4'.\par
\begin{figure}[!htbp]
\includegraphics[width=7.5cm,height=6.5cm]
{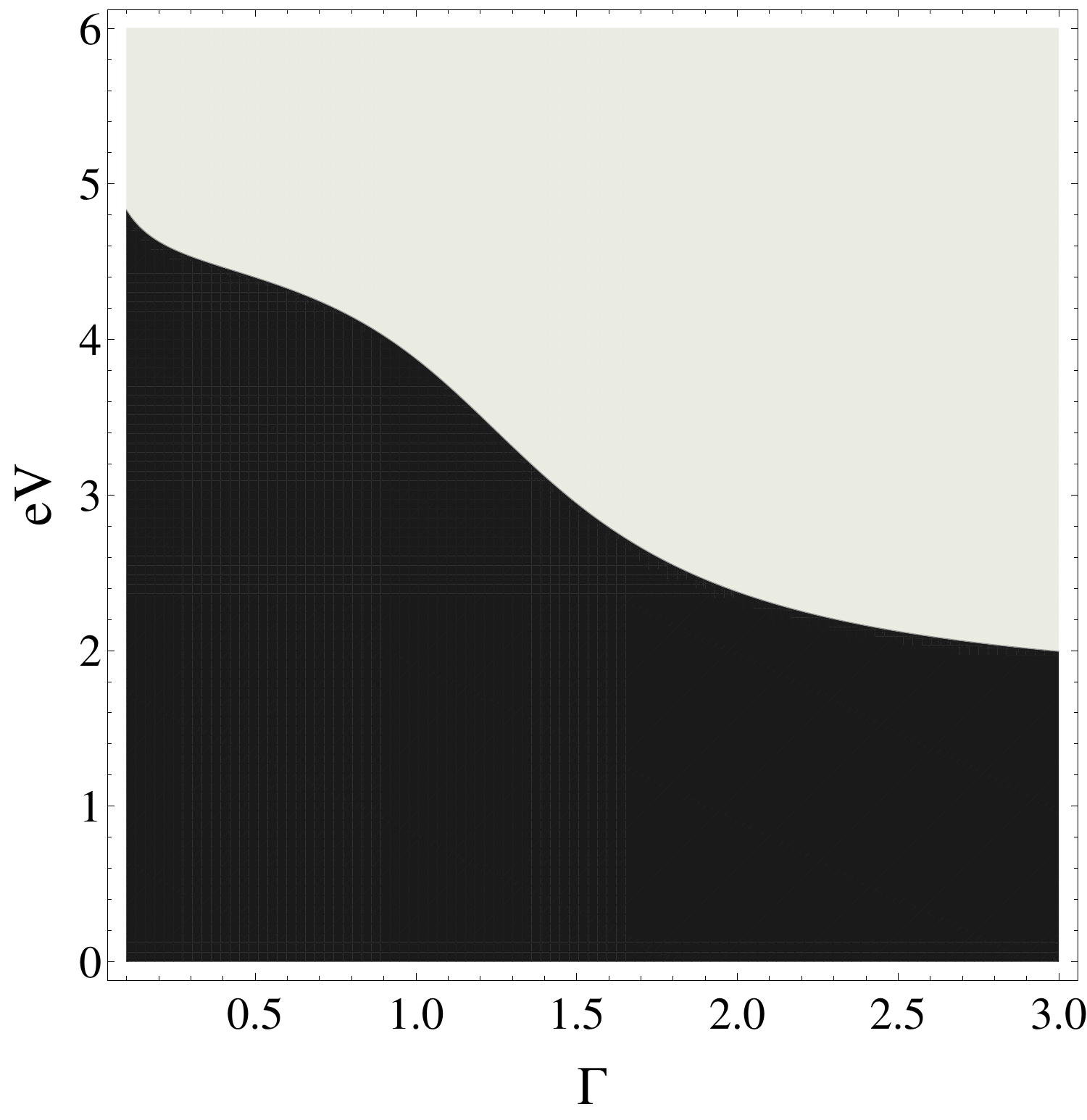}
\caption{Circulating current as a function 
of $\Gamma$ and $eV$ with $\mu_L=\mu+\frac{eV}{2}$, $\mu_R=\mu-\frac{eV}{2}$, 
$\mu=0$, $\beta=100$, $t=1$ and $\epsilon=0$. Here it is shown that 
switching between phases with circulating current (black region) and without circulating 
current (white region) can be done by tuning $\Gamma$ for certain $eV$.}
\label{fig-10}
\end{figure}

\indent
Circulating current appears due to the negativity of $I_{12}$ which comes from 
the negativity of $T_{12}$ in the region 
$|\omega|<\frac{t}{\sqrt{2}}$, outside which $T_{12}$ is positive. For 
$\Gamma\rightarrow\infty$ (specifically $\Gamma \gg t$), $T_{12}$ in the region $|\omega|<\frac{t}{\sqrt{2}}$ goes to zero as 
$T_{12}\approx\frac{(2\omega^2-t^2)}{(\omega^2-t^2)^2}\frac{16}{\Gamma^2}$ and 
hence negative 
contribution to $I_{2\rightarrow1}$ vanishes asymptotically. Therefore for 
large $\Gamma$, circulating current vanishes. 
Thus for sufficiently large bias (with $\mu=0$), greater than $\frac{t}{\sqrt{2}}$, 
and at low temperatures it is be possible to change the sign of 
$I_{12}$ (from negative to positive) by tuning the coupling strength, $\Gamma$, 
(for high temperature this can happen even for $|eV|<\frac{t}{\sqrt{2}}$). 
Hence it is possible to switch between the phases with and without circulating 
currents in the molecule by tuning $\Gamma$. This is shown in Fig. (\ref{fig-10}), 
where the black region represents circulating current in the ring and white region 
represents a region with no circulating current. It 
is clear that, for certain values of $eV$, it is possible to switch between phases 
with and without circulating current by changing $\Gamma$ values. 
Note that $T_{14}$ is also negative over a small energy window, 
$\frac{t}{\sqrt{2}}<|\omega|<t$ which is always compensated by the positive contribution 
from the region $|\omega|<\frac{t}{\sqrt{2}}$, 
leading to positive $I_{4 \rightarrow 1}$. 
\section{Conclusions}
\label{con}
We have studied the bond currents in simple ring shaped molecular junction in 
presence of asymmetry and magnetic field. 
First case studied is symmetric Aharonov-Bohm ring coupled to metal leads, where 
we identified two contributions to 
bond currents, one induced by applied magnetic field ($I_{\phi}$) and the other 
due to applied bias ($I_{V}$). These two contributions have different origins, 
the term $I_{\phi}$ is due to the population terms in eigenstate basis and the 
term $I_{V}$ is due to the coherences induced by leads between different 
eigenstates. It is possible to tune the applied bias and the applied magnetic 
field to completely suppress current across one branch and enhance current 
across the other branch. Lead induced dephasing suppresses the circulating current which, 
for large lead couplings, dies off quadratically ($\approx\frac{1}{\Gamma^2}$). When an 
asymmetry is introduced by coupling a substituent on one of its branches, it is possible 
to generate a circulating current at finite bias, even in the absence of applied magnetic field by tuning 
the coupling strength of the substituent.  
Furthermore, we find that it is possible to switch between phases with and without circulating 
currents by tuning the coupling strength of the molecule with leads. 
\section*{Acknowledgements}
H. Y. and U. H. acknowledge the financial support from the Indian Institute of
Science, Bangalore, India.
\begin{widetext}
\section{Appendix}
\subsection{Eigenstate picture of bond currents in Symmetric Aharonov-Bohm ring}
The isolated molecule in the presence of magnetic flux is described by the 
Hamiltonian expressed in terms of Fock space operators,
\begin{eqnarray}
\label{eq-s1}
  \hat{H}=&\begin{pmatrix}
          c_1^\dag&c_2^\dag&c_3^\dag&c_4^\dag\\
         \end{pmatrix}
         H_{system}
         \begin{pmatrix}
          c_1^{}\\
          c_2^{}\\
          c_3^{}\\
          c_4^{}\\
         \end{pmatrix}
\end{eqnarray}
with $H_{system}$ given by Eq.(\ref{eq-2}). The eigenstates of single particle
Hamiltonian (given by Eq.(\ref{eq-2})) are given by 
\begin{eqnarray}
\label{eq-s2}
\psi_1=\frac{1}{2}\begin{pmatrix}1\\1\\1\\1\end{pmatrix},
\psi_2=\frac{1}{2}\begin{pmatrix}-i\\-1\\i\\1\end{pmatrix}, 
\psi_3=\frac{1}{2}\begin{pmatrix}i\\-1\\-i\\1\end{pmatrix}\text{and }
\psi_4=\frac{1}{2}\begin{pmatrix}-1\\1\\-1\\1\end{pmatrix}
\end{eqnarray}
with corresponding energies 
\begin{eqnarray}
\label{eq-s3}
\epsilon_{1}=-2\cos(\frac{\phi }{4}),
\epsilon_{2}=2\sin(\frac{\phi }{4}),
\epsilon_{3}=-2\sin(\frac{\phi }{4}) \text{and }
\epsilon_{4}=2\cos(\frac{\phi }{4})
\end{eqnarray}
respectively. Hamiltonian in eigenbasis is expressed as
\begin{eqnarray}
\label{eq-s4}
\hat{H}=\sum_{i=1}^{4}\epsilon_i A_i^\dag A_i
\end{eqnarray}
where creation/annihilation ($A_i^\dag \/ A_i^{}$) operators in eigenbasis can be
expressed in terms of creation/annihilation operators in the local basis as,
\begin{eqnarray}
\label{eq-s8}
  \begin{pmatrix}
  A^{}_1\\A^{}_2\\A^{}_3\\A^{}_4
 \end{pmatrix}
 = \mathcal{U}^\dag
  \begin{pmatrix}
  c^{}_1\\c^{}_2\\c^{}_3\\c^{}_4
 \end{pmatrix}
\end{eqnarray}
where matrix $\mathcal{U}$ has single particle eigenstates given in 
Eq.(\ref{eq-s2}) as columns.
Similarly, from Eq. (\ref{eq-4}) and (\ref{eq-5}), $\hat{I}_{2\rightarrow 1}$ and $\hat{I}_{4\rightarrow 1}$ in 
eigenbasis is given by
\begin{eqnarray}
\label{eq-s5}
 \hat{I}_{2\rightarrow 1/4\rightarrow 1}&=&\frac{i}{\hbar}
 \begin{pmatrix}A^{\dag}_1 & A^{\dag}_2 & A^{\dag}_3 & A^{\dag}_4\end{pmatrix} 
 I_{{bond}_{2\rightarrow 1/4\rightarrow 1}}
 \begin{pmatrix} A^{}_1\\A^{}_2\\A^{}_3\\A^{}_4\end{pmatrix}
\end{eqnarray}
where
\begin{eqnarray}
\label{eq-s6}
 I_{{bond}_{2\rightarrow 1}}=\begin{pmatrix}\frac{1}{2} i \sin (\frac{\phi 
}{4}) 
&
(\frac{1}{4}-\frac{i}{4}) (\sin (\frac{\phi}{4})+\cos (\frac{\phi }{4})) &
(\frac{1}{4}+\frac{i}{4}) (\cos (\frac{\phi}{4})-\sin (\frac{\phi }{4})) &
-\frac{1}{2} \cos (\frac{\phi }{4}) \\
 (-\frac{1}{4}-\frac{i}{4}) (\sin (\frac{\phi }{4})+\cos (\frac{\phi }{4})) & 
\frac{1}{2} i \cos (\frac{\phi }{4}) & \frac{1}{2} \sin (\frac{\phi }{4}) & 
(\frac{1}{4}-\frac{i}{4}) (\cos (\frac{\phi }{4})-\sin (\frac{\phi }{4}))  \\
 (-\frac{1}{4}+\frac{i}{4}) (\cos (\frac{\phi }{4})-\sin (\frac{\phi }{4})) & 
-\frac{1}{2} \sin (\frac{\phi }{4}) & -\frac{1}{2} i \cos (\frac{\phi }{4}) &  
(\frac{1}{4}+\frac{i}{4}) (\sin (\frac{\phi }{4})+\cos (\frac{\phi }{4}))  \\
 \frac{1}{2} \cos (\frac{\phi }{4}) & (-\frac{1}{4}-\frac{i}{4}) (\cos
(\frac{\phi}{4})-\sin (\frac{\phi }{4})) & (-\frac{1}{4}+\frac{i}{4}) (\sin
(\frac{\phi }{4})+\cos (\frac{\phi }{4})) & -\frac{1}{2} i \sin
(\frac{\phi}{4})\end{pmatrix}\nonumber\\
\end{eqnarray}
and
\begin{eqnarray}
\label{eq-s7}
 I_{{bond}_{4\rightarrow 1}}=\begin{pmatrix}
 -\frac{1}{2} i \sin (\frac{\phi }{4}) & (-\frac{1}{4}-\frac{i}{4}) (\cos
(\frac{\phi
   }{4})+\sin (\frac{\phi }{4})) & (-\frac{1}{4}+\frac{i}{4}) (\cos
   (\frac{\phi }{4})-\sin (\frac{\phi }{4})) & -\frac{1}{2} \cos (\frac{\phi
}{4})
   \\
 (\frac{1}{4}-\frac{i}{4}) (\cos (\frac{\phi }{4})+\sin (\frac{\phi }{4})) &
   -\frac{1}{2} i \cos (\frac{\phi }{4}) & \frac{1}{2} \sin (\frac{\phi }{4}) &
   (-\frac{1}{4}-\frac{i}{4}) (\cos (\frac{\phi }{4})-\sin (\frac{\phi }{4}))
   \\
 (\frac{1}{4}+\frac{i}{4}) (\cos (\frac{\phi }{4})-\sin (\frac{\phi }{4})) &
   -\frac{1}{2} \sin (\frac{\phi }{4}) & \frac{1}{2} i \cos (\frac{\phi }{4}) &
   (-\frac{1}{4}+\frac{i}{4}) (\cos (\frac{\phi }{4})+\sin (\frac{\phi }{4}))
   \\
 \frac{1}{2} \cos (\frac{\phi }{4}) & (\frac{1}{4}-\frac{i}{4}) (\cos
(\frac{\phi
   }{4})-\sin (\frac{\phi }{4})) & (\frac{1}{4}+\frac{i}{4}) (\cos (\frac{\phi
   }{4})+\sin (\frac{\phi }{4})) & \frac{1}{2} i \sin (\frac{\phi
}{4})\end{pmatrix}.\nonumber\\
\end{eqnarray}
Here diagonal elements of $I_{{bond}_{\alpha \rightarrow 1}}$ multiplied by 
their respective populations give bond currents (between sites '$\alpha$' and 
'1' for $\alpha=2,4$) carried by different eigenstates.\par
\indent
It is clear that in isolated molecule described by a thermal ensemble, 
only populations contribute to bond currents. But if the molecule is
connected to leads, coherences can be induced between eigenstates and hence
bond currents also change.
It is clear, contrary to the recent work \cite{Archak}, eigenstate Lindblad
master equation can give nonzero bond currents (as eigenstates themselves carry 
finite currents), albeit a wrong result out of equilibrium.\par
\indent Lesser Green's function matrix, $G^{<}(\omega)$, can be transformed into 
eigenbasis as, 
$\tilde{G}^{<}(\omega)=\mathcal{U}^{\dag}G^{<}(\omega)\mathcal{U}$. From this, 
bond 
currents are calculated using, $I_{\alpha \rightarrow 
1}=\int^{+\infty}_{-\infty}\frac{d\omega}{2\pi}Tr[I_{{bond}_{\alpha \rightarrow 
1}}\tilde{G}^{<}(\omega)]$ for $\alpha=2,4$. By explicit 
calculation it can be seen (for the case $\Gamma_L=\Gamma_R=\Gamma$) that, 
$I_{2\rightarrow 1}=I_{V}+I_{\phi}$ and $I_{4\rightarrow 1}=I_{V}-I_{\phi}$, 
where only population terms of $\tilde{G}^{<}(\omega)$ contribute to 
$I_{\phi}$ and coherences contribute to $I_{V}$.\par
\indent For thermodynamic equilibrium (i.e., $\mu_L=\mu_R=\mu$ and 
$\beta_L=\beta_R=\beta$), eigenstate energies, eigenstate populations 
($-i\int_{-\infty}^{+\infty}\frac{d\omega}{2\pi}G_{mm}^{<}(\omega)$) and 
eigenstate contributions to $I_{\phi}$ are periodic in $\phi$ with period 
$8\pi$ as shown in Figs. (\ref{fig-s1}), (\ref{fig-s2}) and (\ref{fig-s3}). The 
net contribution of each eigenstate is also periodic in $\phi$ with period 
$8\pi$ as shown in Fig. (\ref{fig-s4}), but $I_{\phi}$ is periodic in $\phi$ 
with period $2\pi$ as can be seen in Fig. (\ref{fig-s5}). 
This is because eigenstate energies, eigenstate contributions ($I_{{bond}_{{2 
\rightarrow 1}_{mm}}}$) to $I_\phi$ and eigenstate populations get swapped 
after $2\pi$ increment in $\phi$ as can be seen in Figs. (\ref{fig-s1}), 
(\ref{fig-s2}) and (\ref{fig-s3}). Furthermore, for $\phi=\pi$, 
states with opposites contributions to current $I_{\phi}$ become degenerate, hence 
the circulating current vanishes.
\begin{figure}[!htbp]
           \centering
            \begin{minipage}[b]{0.4\textwidth}
             \includegraphics[width=7.5cm,height=5.5cm]{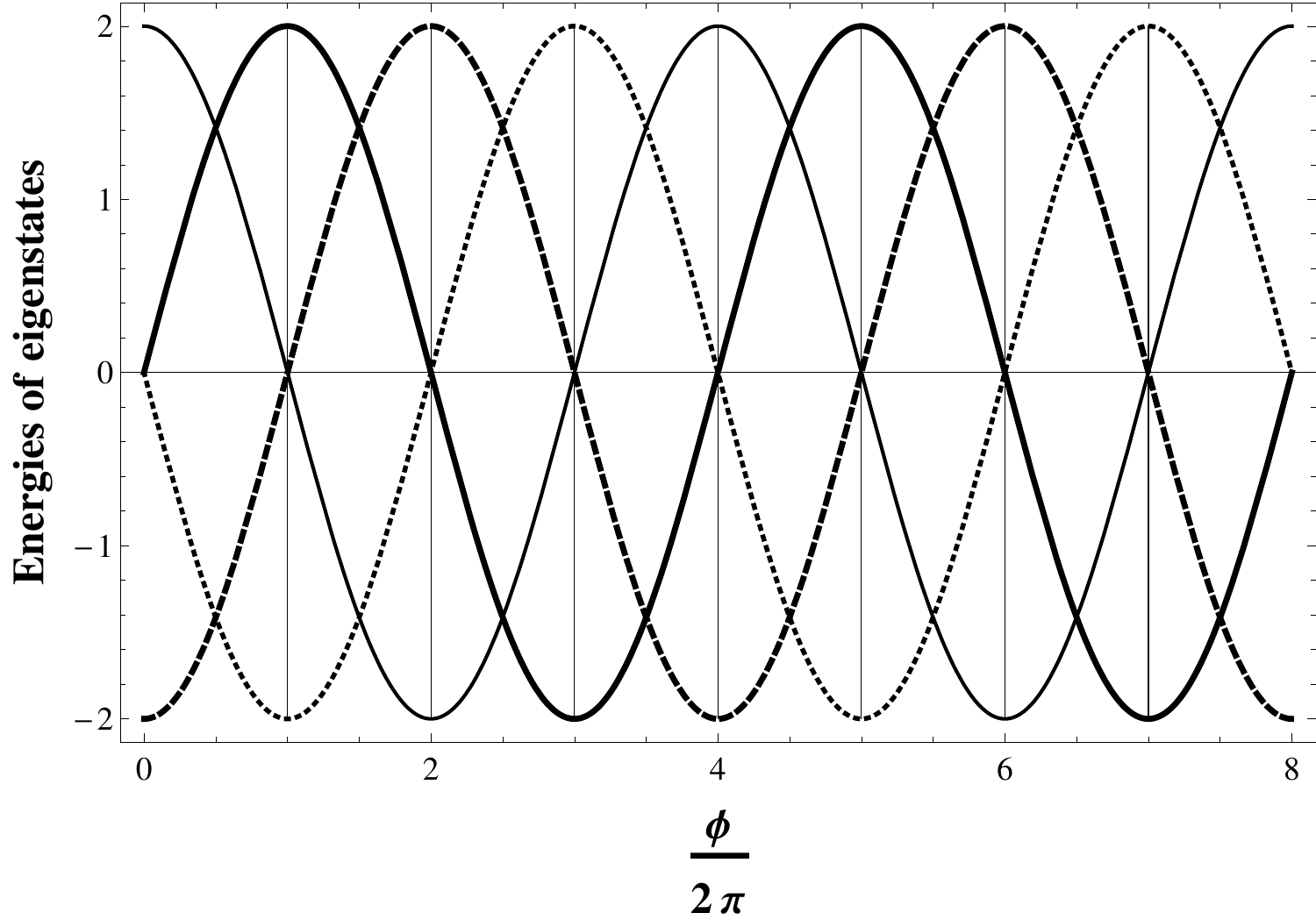}
             \caption{Eigenvalues of isolated ring as a function of $\phi$. Dashed, thick, dotted and thin curves represent eigenstate energies of states 1, 2, 3 and 4.}
             \label{fig-s1}
             \end{minipage}
             \hfill
             \begin{minipage}[b]{0.4\textwidth}
             \includegraphics[width=7.5cm,height=5.5cm]{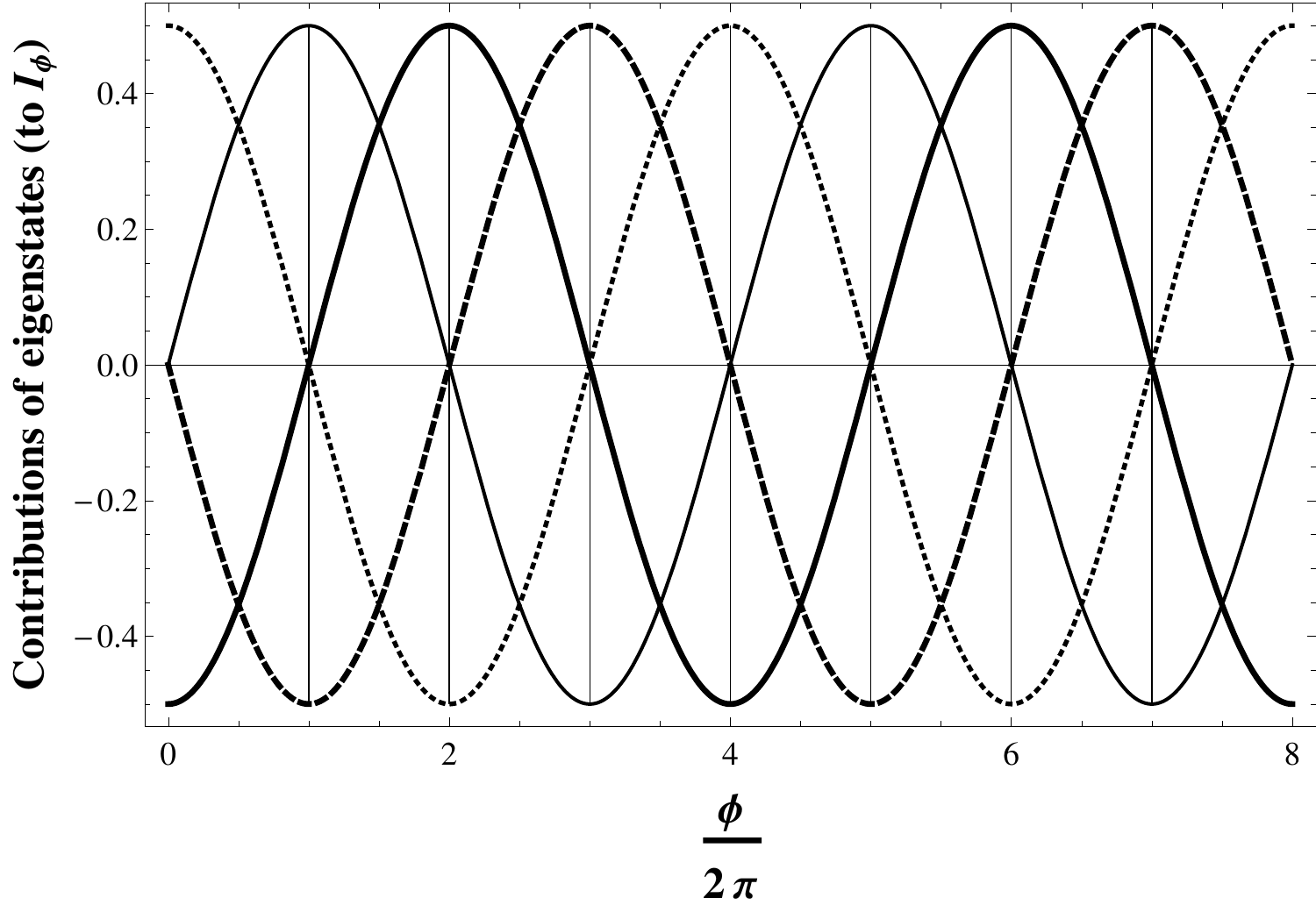}
             \caption{Contribution of various eigenstates to $I_{\phi}$ as a function of $\phi$. Dashed, thick, dotted and thin curves represent contribution of eigenstates 1, 2, 3 and 4 to $I_{\phi}$.}
             \label{fig-s2}
             \end{minipage}
             \hfill
             \begin{minipage}[b]{0.4\textwidth}
             \includegraphics[width=7.5cm,height=5.5cm]{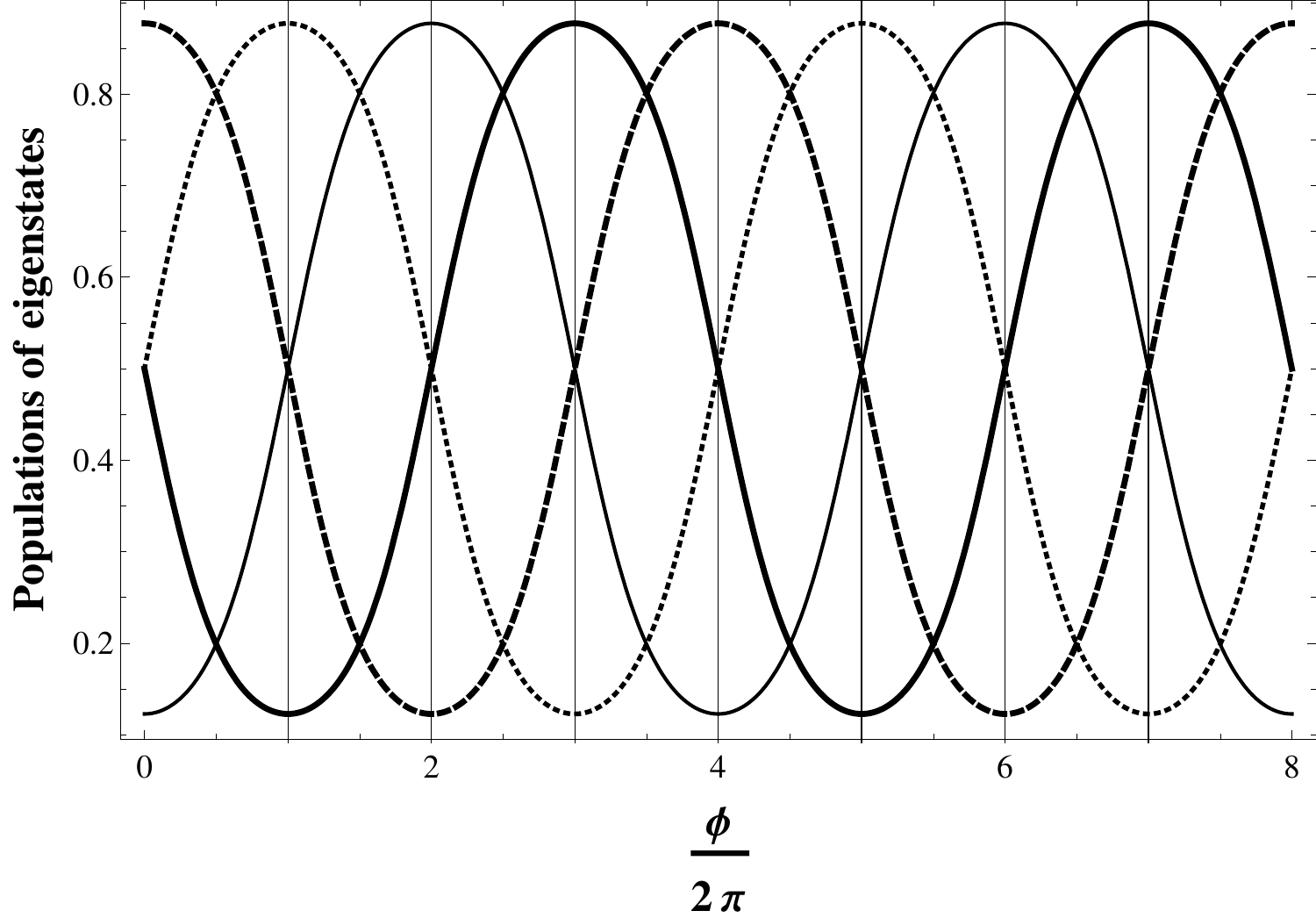}
             \caption{Populations of various eigenstates of ring connected to reservoir as a function of $\phi$ with $\mu=0$, $eV=0$, $\beta=1$ and $\Gamma=0.1$. Dashed, thick, dotted and thin curves represent populations of eigenstates 1, 2, 3 and 4.}
             \label{fig-s3}
             \end{minipage}
             \hfill
             \begin{minipage}[b]{0.4\textwidth}
             \includegraphics[width=7.5cm,height=5.5cm]{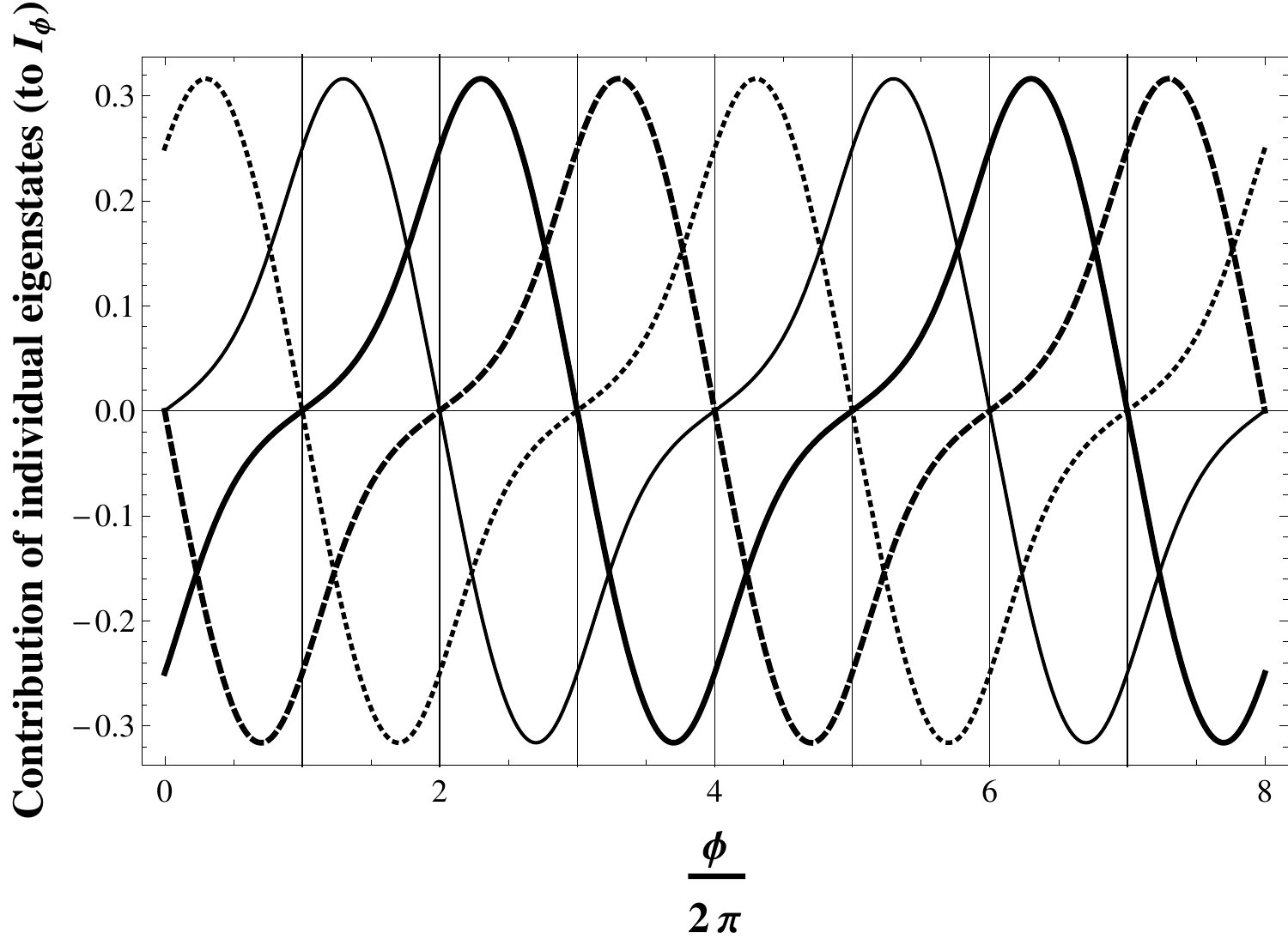}
             \caption{Individual contributions of eigenstates as a function of $\phi$ with $\mu=0$, $eV=0$, $\beta=1$ and $\Gamma=0.1$. Dashed, thick, dotted and thin curves represent individual contributions of eigenstates 1, 2, 3 and 4 to $I_{\phi}$.}
             \label{fig-s4}
             \end{minipage}
\end{figure}  
\begin{figure}[!htbp]
\includegraphics[width=7.5cm,height=5.5cm]{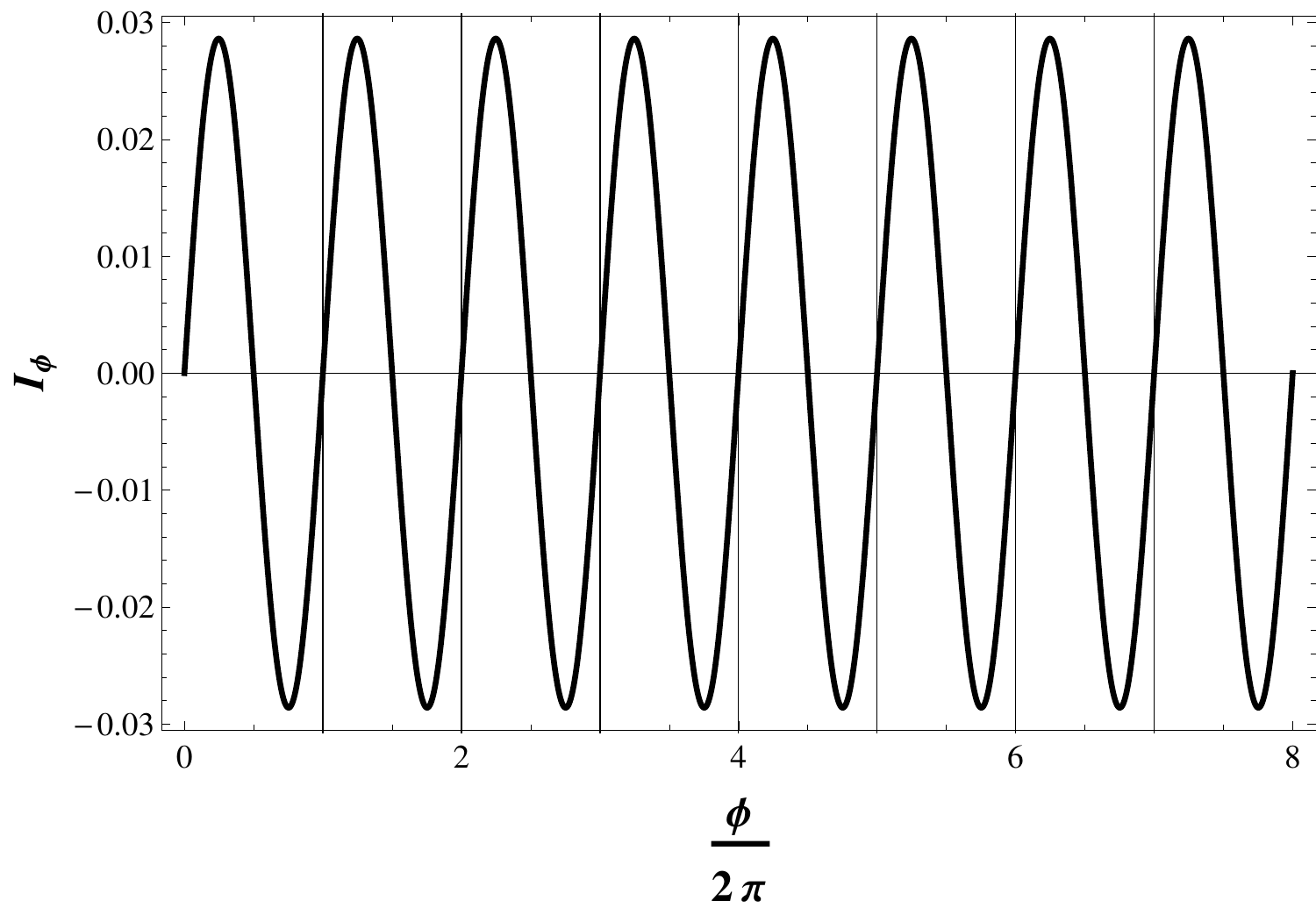}
\caption{$I_{\phi}$ as a function of $\phi$ with $\mu=0$, $eV=0$, $\beta=1$ and 
$\Gamma=0.1$.}
\label{fig-s5}
\end{figure}
\subsection{Spatial path picture of net current in symmetric Aharonov-Bohm ring}
For non interacting electron systems considered here, the net current in the 
circuit (which can be obtained by using Eq. (\ref{eq-14}) in Eq. (\ref{eq-8})) can be expressed as, 
\begin{eqnarray}
\label{eq-s9}
 I_{L}&=&\int_{-\infty}^{+\infty}\frac{d\omega}{2\pi}[\Gamma_L 
G^{r}_{13}(\omega)\Gamma_R G^{a}_{31}(\omega)][f_L(\omega)-f_R(\omega)].
\end{eqnarray}
Following Ref. \cite{Hansen}, we use projection operator technique to project out sites '2' 
and '4' and obtain, 
\begin{eqnarray}
\label{eq-s10}
&&G^{r}_{13}(\omega)=\frac{1}{\omega+i\frac{\Gamma_L}{2}-\Sigma_{11}^{Upper}
-\Sigma_{11}^{Lower}-\frac{(\Sigma_{13}^{Upper}+\Sigma_{13}^{Lower})(\Sigma_{31}
^{Upper}+\Sigma_{31}^{Lower})}{\omega+i\frac{\Gamma_R}{2}-\Sigma_{33}^{Upper}
-\Sigma_{33}^{Lower}}}\times(\Sigma_{13}^{Upper}+\Sigma_{13}^{Lower})
 \times 
\frac{1}{\omega+i\frac{\Gamma_R}{2}-\Sigma_{33}^{Upper}-\Sigma_{33}^{Lower}}.
 \end{eqnarray}
The first term in the product corresponds to renormalized retarded Green's 
function for site '1' with self energies coming from excursions into upper 
branch ($\Sigma_{11}^{Upper}$), lower branch ($\Sigma_{11}^{Lower}$) and to and 
fro excursions to site '3' 
($\frac{(\Sigma_{13}^{Upper}+\Sigma_{13}^{Lower})(\Sigma_{31}^{Upper}+\Sigma_{31
}^{Lower})}{\omega+i\frac{\Gamma_R}{2}-\Sigma_{33}^{Upper}-\Sigma_{33}^{Lower}}
$). 
The second term corresponds to the sum of bare amplitudes to go from site '1' 
to 
site '3' through upper ($\Sigma_{13}^{Upper}$) and lower branches 
($\Sigma_{13}^{Lower}$). 
The third term corresponds to the retarded Green's function for site '3' with 
self energies coming from excursions into upper branch ($\Sigma_{33}^{Upper}$) 
and lower branch ($\Sigma_{33}^{Lower}$) only. 
$\Sigma^{Upper/Lower}_{ab}$ are matrix elements of self energies due to upper 
or 
lower branches respectively and they are given by 
$\Sigma^{Upper}=\frac{1}{\omega}\begin{pmatrix}1 && 
e^{i\frac{\phi}{2}}\\e^{-i\frac{\phi}{2}} && 1 \end{pmatrix}$ and 
$\Sigma^{Lower}=\frac{1}{\omega}\begin{pmatrix}1 && 
e^{-i\frac{\phi}{2}}\\e^{i\frac{\phi}{2}} && 1 \end{pmatrix}$. 
$G^{a}_{31}(\omega)$ can be obtained as 
$G^{a}_{31}(\omega)=(G^{r}_{13}(\omega))^{*}|_{\phi \rightarrow -\phi}$. For 
$\phi=\pi$, the two pathways $1\rightarrow 2\rightarrow 3$ and $1\rightarrow 
4\rightarrow 3$ destructively interfere (since 
$(\Sigma_{13}^{Upper}+\Sigma_{13}^{Lower})|_{\phi=\pi}=(\frac{e^{i\frac{\phi}{2}
}}{\omega}+\frac{e^{-i\frac{\phi}{2}}}{\omega})|_{\phi=\pi}=0$) leading to zero 
net current in the circuit. 
Also for $\phi\neq2n\pi$ ('n' is any integer), the net transmission 
function $T_{L}(\omega)=\Gamma_L G^{r}_{13}(\omega)\Gamma_R G^{a}_{31}(\omega)$ 
has a zero (anti-resonance) at $\omega=0$, which is a resonance zero 
\cite{Hansen} (since the particle injected from lead into the system at this 
energy 
is in resonance with sites '2' and '4').
\subsection{Analytical expressions for currents for the Symmetric Aharonov-Bohm ring}
\subsubsection*{Finite temperature expressions}
$\omega$ integrals in Eqs. (\ref{eq-18}) and (\ref{eq-19}) can be performed using contour integration technique to get,
\begin{eqnarray}
\label{eq-s11}
 I_{V}&=&\frac{\Gamma^2\cos^2(\frac{\phi}{2})}{2\pi}\times\nonumber\\
 &&\bigg[\frac{i a_1}{(a_1^2-a_2^2)(a_1^2-a_3^2)(a_1^2-a_4^2)}\times\nonumber\\
&&\Big\{\Psi[\frac{1}{2}-i\frac{\beta}{2\pi}(\mu_L+ia_1)]-\Psi[\frac{1}{2}+i\frac{\beta}{2\pi}(\mu_L-ia_1)]
+\Psi[\frac{1}{2}+i\frac{\beta}{2\pi}(\mu_R-ia_1)]-\Psi[\frac{1}{2}-i\frac{\beta}{2\pi}(\mu_R+ia_1)]\Big\}\nonumber\\
 &+&\frac{i a_2}{(a_2^2-a_1^2)(a_2^2-a_3^2)(a_2^2-a_4^2)}\times\nonumber\\
&&\Big\{\Psi[\frac{1}{2}-i\frac{\beta}{2\pi}(\mu_L+ia_2)]-\Psi[\frac{1}{2}+i\frac{\beta}{2\pi}(\mu_L-ia_2)]
+\Psi[\frac{1}{2}+i\frac{\beta}{2\pi}(\mu_R-ia_2)]-\Psi[\frac{1}{2}-i\frac{\beta}{2\pi}(\mu_R+ia_2)]\Big\}\nonumber\\
 &+&\frac{i a_3}{(a_3^2-a_1^2)(a_3^2-a_2^2)(a_3^2-a_4^2)}\times\nonumber\\
&&\Big\{\Psi[\frac{1}{2}-i\frac{\beta}{2\pi}(\mu_L+ia_3)]-\Psi[\frac{1}{2}+i\frac{\beta}{2\pi}(\mu_L-ia_3)]
+\Psi[\frac{1}{2}+i\frac{\beta}{2\pi}(\mu_R-ia_3)]-\Psi[\frac{1}{2}-i\frac{\beta}{2\pi}(\mu_R+ia_3)]\Big\}\nonumber\\
 &+&\frac{i a_4}{(a_4^2-a_1^2)(a_4^2-a_2^2)(a_4^2-a_3^2)}\times\nonumber\\
&&\Big\{\Psi[\frac{1}{2}-i\frac{\beta}{2\pi}(\mu_L+ia_4)]-\Psi[\frac{1}{2}+i\frac{\beta}{2\pi}(\mu_L-ia_4)]
+\Psi[\frac{1}{2}+i\frac{\beta}{2\pi}(\mu_R-ia_4)]-\Psi[\frac{1}{2}-i\frac{\beta}{2\pi}(\mu_R+ia_4)]\Big\}\bigg]\nonumber\\
\end{eqnarray}
and
\begin{eqnarray}
\label{eq-s12}
 I_{\phi}&=&\frac{\Gamma\sin(\phi)}{2\pi}\times\nonumber\\
&&\bigg[\frac{a_1^2+2}{(a_1^2-a_2^2)(a_1^2-a_3^2)(a_1^2-a_4^2)}\times\nonumber\\
&&\Big\{\Psi[\frac{1}{2}+i\frac{\beta}{2\pi}(\mu_L-ia_1)]+\Psi[\frac{1}{2}-i\frac{\beta}{2\pi}(\mu_L+ia_1)]
+\Psi[\frac{1}{2}+i\frac{\beta}{2\pi}(\mu_R-ia_1)]+\Psi[\frac{1}{2}-i\frac{\beta}{2\pi}(\mu_R+ia_1)]\Big\}\nonumber\\
&+&\frac{a_2^2+2}{(a_2^2-a_1^2)(a_2^2-a_3^2)(a_2^2-a_4^2)}\times\nonumber\\
&&\Big\{\Psi[\frac{1}{2}+i\frac{\beta}{2\pi}(\mu_L-ia_2)]+\Psi[\frac{1}{2}-i\frac{\beta}{2\pi}(\mu_L+ia_2)]
+\Psi[\frac{1}{2}+i\frac{\beta}{2\pi}(\mu_R-ia_2)]+\Psi[\frac{1}{2}-i\frac{\beta}{2\pi}(\mu_R+ia_2)]\Big\}\nonumber\\
&+&\frac{a_3^2+2}{(a_3^2-a_1^2)(a_3^2-a_2^2)(a_3^2-a_4^2)}\times\nonumber\\
&&\Big\{\Psi[\frac{1}{2}+i\frac{\beta}{2\pi}(\mu_L-ia_3)]+\Psi[\frac{1}{2}-i\frac{\beta}{2\pi}(\mu_L+ia_3)]
+\Psi[\frac{1}{2}+i\frac{\beta}{2\pi}(\mu_R-ia_3)]+\Psi[\frac{1}{2}-i\frac{\beta}{2\pi}(\mu_R+ia_3)])\Big\}\nonumber\\
&+&\frac{a_4^2+2}{(a_4^2-a_1^2)(a_4^2-a_2^2)(a_4^2-a_3^2)}\times\nonumber\\
&&\Big\{\Psi[\frac{1}{2}+i\frac{\beta}{2\pi}(\mu_L-ia_4)]+\Psi[\frac{1}{2}-i\frac{\beta}{2\pi}(\mu_L+ia_4)]
+\Psi[\frac{1}{2}+i\frac{\beta}{2\pi}(\mu_R-ia_4)]+\Psi[\frac{1}{2}-i\frac{\beta}{2\pi}(\mu_R+ia_4)]\Big\}\bigg],\nonumber\\
\end{eqnarray}
where $a_1=\frac{\Gamma+\sqrt{\Gamma^2-64\sin^2(\frac{\phi}{4})}}{4}$,
$a_2=\frac{\Gamma-\sqrt{\Gamma^2-64\sin^2(\frac{\phi}{4})}}{4}$,
$a_3=\frac{\Gamma+\sqrt{\Gamma^2-64\cos^2(\frac{\phi}{4})}}{4}$,
$a_4=\frac{\Gamma-\sqrt{\Gamma^2-64\cos^2(\frac{\phi}{4})}}{4}$ and $\Psi[z]$ 
is digamma function in variable 'z' \cite{Thomson}.
\subsubsection*{Zero temperature expressions}
$\omega$ integrals can be performed after taking zero temperature ($\beta\to\infty$) limits in Eqs. (\ref{eq-18}) and (\ref{eq-19}) to get,
\begin{eqnarray}
\label{eq-s13}
 I_{V}&=&2\Gamma^2\cos^2(\frac{\phi}{2})\times\nonumber\\
 &&\bigg[\frac{a_1}{(a_1^2-a_2^2)(a_1^2-a_3^2)(a_1^2-a_4^2)}\big[\arctan(\frac{
\mu_L}{a_1})-\arctan(\frac{\mu_R}{a_1})\big]
 +\frac{a_2}{(a_2^2-a_1^2)(a_2^2-a_3^2)(a_2^2-a_4^2)}\big[\arctan(\frac{\mu_L}{
a_2})-\arctan(\frac{\mu_R}{a_2})\big]\nonumber\\
 &+&\frac{a_3}{(a_3^2-a_1^2)(a_3^2-a_2^2)(a_3^2-a_4^2)}\big[\arctan(\frac{\mu_L}
{a_3})-\arctan(\frac{\mu_R}{a_3})\big]
 +\frac{a_4}{(a_4^2-a_1^2)(a_4^2-a_2^2)(a_4^2-a_3^2)}\big[\arctan(\frac{\mu_L}{
a_4})-\arctan(\frac{\mu_R}{a_4})\big]\bigg]\nonumber\\
\end{eqnarray}
and
\begin{eqnarray}
\label{eq-s14}
 I_{\phi}&=&2\Gamma\sin(\phi)\times\nonumber\\
 &&\bigg[\frac{a_1^2+2}{(a_1^2-a_2^2)(a_1^2-a_3^2)(a_1^2-a_4^2)}\big[
\ln(\mu_L^2+a_1^2)+\ln(\mu_R^2+a_1^2)\big]
 +\frac{a_2^2+2}{(a_2^2-a_1^2)(a_2^2-a_3^2)(a_2^2-a_4^2)}\big[
\ln(\mu_L^2+a_2^2)+\ln(\mu_R^2+a_2^2)\big]\nonumber\\
 &+&\frac{a_3^2+2}{(a_3^2-a_1^2)(a_3^2-a_2^2)(a_3^2-a_4^2)}\big[
\ln(\mu_L^2+a_3^2)+\ln(\mu_R^2+a_3^2)\big]
 +\frac{a_4^2+2}{(a_4^2-a_1^2)(a_4^2-a_2^2)(a_4^2-a_3^2)}\big[
\ln(\mu_L^2+a_4^2)+\ln(\mu_R^2+a_4^2)\big]\bigg].\nonumber\\
\end{eqnarray}
\subsection{Interpretation of zeros of Transmission functions for asymmetric 
ring junction}
Using Eq. (\ref{eq-14}) in Eq. (\ref{eq-8}), the net current in the circuit can 
be recast as, 
\begin{eqnarray}
\label{eq-26}
 &&I_{L}=\nonumber\\
 &&\int_{-\infty}^{+\infty}\frac{d\omega}{2\pi}[\Gamma_L G^{r}_{13}(\omega) 
\Gamma_R G^{a}_{31}(\omega)][f_L(\omega)-f_R(\omega)].
\end{eqnarray}
Similar to Appendix B, we project out sites '2', '4' and '5' to obtain expression for 
$G^{r}_{13}(\omega)$ as,
\begin{eqnarray}
\label{eq-27}
&&G^{r}_{13}(\omega)=\frac{1}{\omega+i\frac{\Gamma_L}{2}-\Sigma_{11}^{Upper}
-\Sigma_{11}^{Lower}-\frac{(\Sigma_{13}^{Upper}+\Sigma_{13}^{Lower})(\Sigma_{31}
^{Upper}+\Sigma_{31}^{Lower})}{\omega+i\frac{\Gamma_R}{2}-\Sigma_{33}^{Upper}
-\Sigma_{33}^{Lower}}}\times(\Sigma_{13}^{Upper}+\Sigma_{13}^{Lower})
 \times 
\frac{1}{\omega+i\frac{\Gamma_R}{2}-\Sigma_{33}^{Upper}-\Sigma_{33}^{Lower}},
\end{eqnarray}
The interpretation of three terms in $G^{a}_{13}(\omega)$ is same as discussed 
in Appendix B.
$\Sigma^{Upper/Lower}_{ab}$ are matrix elements of self energies due to upper 
(lower) branch, they are given by 
$\Sigma^{Upper}=\frac{(\omega-\epsilon)}{\omega(\omega-\epsilon)-t^2}
\begin{pmatrix}1 && 1 \\ 1 && 1 \end{pmatrix}$ and 
$\Sigma^{Lower}=\frac{1}{\omega}\begin{pmatrix} 1 && 1 \\ 1 && 1 
\end{pmatrix}$. 
$G^{a}_{31}(\omega)$ can be obtained as 
$G^{a}_{31}(\omega)=(G^{r}_{13}(\omega))^{*}$. For 
$\omega=\frac{\epsilon\pm\sqrt{\epsilon^2+2t^2}}{2}$, the two pathways 
$1\rightarrow 2\rightarrow 3$ and $1\rightarrow 4\rightarrow 3$ destructively 
interfere (since 
$(\Sigma_{13}^{Upper}+\Sigma_{13}^{Lower})|_{\omega=\frac{\epsilon\pm\sqrt{
\epsilon^2+2t^2}}{2}}=0$) 
leading to zeros (anti-resonances) in the net transmission coefficient (termed 
as multi-path zeros due to their origin from destructive interference between 
the two paths) \cite{Hansen}. 
However in the presence of applied magnetic field, these anti resonances in 
$T_L(\omega)$ disappear (Eq. (\ref{eq-17})).\par
\indent
Using Eq. (\ref{eq-14}) in Eqs. (\ref{eq-6}) and (\ref{eq-7}) (additionally using 
$Im[G^{r}_{13}(\omega)G^{a}_{32}(\omega)]=-Im[G^{r}_{11}(\omega)G^{a}_{12}
(\omega)]$ and 
$Im[G^{r}_{13}(\omega)G^{a}_{34}(\omega)]=-Im[G^{r}_{11}(\omega)G^{a}_{14}
(\omega)]$ for $\phi=0$ case), expressions for the two bond currents 
$I_{2\rightarrow1}$ and $I_{4\rightarrow1}$, given by Eqs. (\ref{eq-6}) and 
(\ref{eq-7}) with $\phi=0$ and 
$\Gamma_L=\Gamma_R=\Gamma$ can be cast as, 
\begin{eqnarray}
\label{eq-28}
 &&I_{2\rightarrow1}=\nonumber\\ 
 &&-2 \int_{-\infty}^{+\infty}\frac{d\omega}{2\pi}\Gamma 
Im[G^{r}_{11}(\omega)G^{a}_{12}(\omega)][f_L(\omega)-f_R(\omega)],
\end{eqnarray}
\begin{eqnarray}
\label{eq-29}
 &&I_{4\rightarrow1}=\nonumber\\ 
 &&-2 \int_{-\infty}^{+\infty}\frac{d\omega}{2\pi}\Gamma 
Im[G^{r}_{11}(\omega)G^{a}_{14}(\omega)][f_L(\omega)-f_R(\omega)].
\end{eqnarray}
\par
\indent To analyze zeros of transmission function $T_{12}(\omega)=-2\Gamma 
Im[G^{r}_{11}(\omega)G^{a}_{12}(\omega)]$, for the current between sites '1' 
and '2', we follow the same procedure as above and project out sites '3', '4' 
and '5' to get, 
\begin{eqnarray}
&&G^{r}_{11}(\omega)=\frac{1}{\omega+i\frac{\Gamma}{2}-\Sigma^{Indirect}_{11}
-\frac{(\Sigma^{Direct}_{12}+\Sigma^{Indirect}_{12})(\Sigma^{Direct}_{21}
+\Sigma^{Indirect}_{21})}{\omega-\Sigma^{Indirect}_{22}-\Sigma^{Substituent}_{22
}}}
\end{eqnarray}
and
\begin{eqnarray}
&&G^{a}_{12}(\omega)=\frac{1}{\omega-i\frac{\Gamma}{2}-(\Sigma^{Indirect}_{11}
)^*-\frac{((\Sigma^{Direct}_{12})^*+(\Sigma^{Indirect}_{12})^*)((\Sigma^{Direct}
_{21})^*+(\Sigma^{Indirect}_{21})^*)}{\omega-(\Sigma^{Indirect}_{22})^*-(\Sigma^
{Substituent}_{22})^*}}\nonumber\\
&&\times((\Sigma^{Direct}_{12})^*+(\Sigma^{Indirect}_{12})^*)\frac{1}{
\omega-(\Sigma^{Indirect}_{22})^*-(\Sigma^{Substituent}_{22})^*}.
\end{eqnarray}
$\Sigma^{Direct/Indirect}_{ab}$ are matrix elements of self energies due to 
direct path '$1 \rightarrow 2$' (indirect path '$1 \rightarrow 4 \rightarrow 3 
\rightarrow 2$') given by $\Sigma^{Direct}=\begin{pmatrix} 0 && -1\\ -1 && 0 
\end{pmatrix}$ and 
$\Sigma^{Direct}=\begin{pmatrix} \frac{1}{\omega - 
\frac{1}{\omega+i\frac{\Gamma}{2}}} && 
-\frac{1}{\omega}\frac{1}{\omega+i\frac{\Gamma}{2}-\frac{1}{\omega}}\\ 
-\frac{1}{\omega+i\frac{\Gamma}{2}-\frac{1}{\omega}}\frac{1}{\omega} && 
\frac{1}{\omega+i\frac{\Gamma}{2} - \frac{1}{\omega}} \end{pmatrix}$, self 
energy due to extra substituent is $\Sigma^{Substituent}=\begin{pmatrix}0 && 0\\ 
0 && \frac{t^2}{\omega-\epsilon} \end{pmatrix}$. 
At $\omega=\epsilon$, $G^{a}_{12}(\omega)$ becomes zero (due to the bare 
advanced Green's function term becoming zero due to divergence of 
$\Sigma_{22}^{Substituent}$), leading to zero of the transmission function 
(termed as resonance zero). Another zero (multi-path zero) of the transmission 
function can be identified at $\omega=0$, where 
$(\Sigma^{Direct}_{12})^*+(\Sigma^{Inirect}_{12})^*=-\frac{\omega}{
\omega(\omega-i\frac{\Gamma}{2})-1}$ becomes zero, which can be interpreted as 
a result of destructive interference between direct and 
indirect paths. Another set of zeros (which are multi-path zeros of net 
transmission function $T_L(\omega)$ at 
$\omega=\frac{\epsilon\pm\sqrt{\epsilon^2+2t^2}}{2}$ discussed above) does not 
have a simple interpretation in this procedure.\par
\indent For analyzing zeros of transmission function $T_{14}(\omega)=-2\Gamma 
Im[G^{r}_{11}(\omega)G^{a}_{14}(\omega)]$, for the current between sites '1' 
and '4', we project out sites '2', '3' and '5' to get,
\begin{eqnarray}
G^{r}_{11}(\omega)=\frac{1}{\omega+i\frac{\Gamma}{2}-\Sigma^{Indirect}_{11}
-\frac{(\Sigma^{Direct}_{14}+\Sigma^{Indirect}_{14})(\Sigma^{Direct}_{41}
+\Sigma^{Indirect}_{41})}{\omega-\Sigma^{Indirect}_{44}}}
\end{eqnarray}
and
\begin{eqnarray}
&&G^{a}_{14}(\omega)=\frac{1}{\omega-i\frac{\Gamma}{2}-(\Sigma^{Indirect}_{11}
)^*-\frac{((\Sigma^{Direct}_{14})^*+(\Sigma^{Indirect}_{14})^*)((\Sigma^{Direct}
_{41})^*+(\Sigma^{Indirect}_{41})^*)}{\omega-(\Sigma^{Indirect}_{44})^*}}
\nonumber\\
&&\times((\Sigma^{Direct}_{14})^*+(\Sigma^{Indirect}_{14})^*)\frac{1}{
\omega-(\Sigma^{Indirect}_{44})^*}.
\end{eqnarray}
$\Sigma^{Direct/Indirect}_{ab}$ are matrix elements of self energies due to 
direct path '$1 \rightarrow 4$' (indirect path '$1 \rightarrow 2 \rightarrow 3 
\rightarrow 4$') given by $\Sigma^{Direct}=\begin{pmatrix} 0 && -1\\ -1 && 0 
\end{pmatrix}$ and 
$\Sigma^{Direct}=\frac{1}{[(\omega+i\frac{\Gamma}{2})\{
\omega(\omega-\epsilon)-t^2\}-(\omega-\epsilon)]}\begin{pmatrix} 
(\omega-\epsilon)(\omega+i\frac{\Gamma}{2}) && -(\omega-\epsilon)\\ 
-(\omega-\epsilon) && \omega(\omega-\epsilon)-t^2 \end{pmatrix}$. 
At $\omega=\frac{\epsilon\pm\sqrt{\epsilon^2+4 t^2}}{2}$,  
$(\Sigma^{Direct}_{14})^*+(\Sigma^{Inirect}_{14})^*=\frac{(\omega+i\frac{\Gamma}
{2})\{\omega(\omega-\epsilon)-t^2\}}{[(\omega+i\frac{\Gamma}{2})\{
\omega(\omega-\epsilon)-t^2\}-(\omega-\epsilon)]}$ becomes zero, hence 
$\omega=\frac{\epsilon\pm\sqrt{\epsilon^2+4 t^2}}{2}$ are zeros of 
$T_{14}(\omega)$ 
(these zeros are a result of destructive interference between direct and 
indirect paths and hence can be termed as multi-path zeros). 
Similar to $T_{12}(\omega)$ case, another set of zeros (at 
$\omega=\frac{\epsilon\pm\sqrt{\epsilon^2+2t^2}}{2}$) does not have a simple 
interpretation in this procedure.
\end{widetext}
\bibliography{nonequilibrium_ring_currents}
\bibliographystyle{unsrt}
\end{document}